\documentclass[fleqn,usenatbib]{mnras}

\usepackage{newtxtext,newtxmath}

\usepackage[T1]{fontenc}

\DeclareRobustCommand{\VAN}[3]{#2}
\let\VANthebibliography\thebibliography
\def\thebibliography{\DeclareRobustCommand{\VAN}[3]{##3}\VANthebibliography}

\usepackage{graphicx}	
\usepackage{amsmath}	
\usepackage{newtxmath}  

\title[A `faint and slow' TDE]{eRASSt~J074426.3+291606: prompt accretion disc formation in a `faint and slow' tidal disruption event}

\author[A. Malyali et al.]{A. Malyali$^{1}$\thanks{E-mail: amalyali@mpe.mpg.de}, Z. Liu$^{1}$, A. Merloni$^{1}$, A.Rau$^{1}$, J. Buchner$^{1}$, S. Ciroi$^{2}$, F. Di Mille$^{3}$, I. Grotova$^{1}$, T. Dwelly$^{1}$,  \newauthor K. Nandra$^{1}$, M. Salvato$^{1}$, D. Homan$^{4}$, M. Krumpe$^{4}$
\\
$^{1}$Max-Planck-Institut f\"ur extraterrestrische Physik,  Giessenbachstrasse 1, 85748 Garching, Germany\\ 
$^{2}$Dipartimento di Fisica e Astronomia, Universit\`a degli Studi di Padova, Vicolo dell'Osservatorio 3, 35122 Padova, Italy\\
$^{3}$Las Campanas Observatory - Carnegie Institution for Science, Colina el Pino, Casilla 601, La Serena, Chile\\
$^{4}$Leibniz-Institut für Astrophysik Potsdam, An der Sternwarte 16, 14482 Potsdam, Germany\\
}

\begin{document}
\label{firstpage}
\pagerange{\pageref{firstpage}--\pageref{lastpage}}
\maketitle

\begin{abstract}
We report on multi-wavelength observations of the tidal disruption event (TDE) candidate eRASSt~J074426.3+291606 (J0744), located in the nucleus of a previously quiescent galaxy at $z=0.0396$. J0744 was first detected as a new, ultra-soft X-ray source (photon index $\sim 4$) during the second \textit{SRG}/eROSITA All-Sky Survey (eRASS2), where it had brightened in the 0.3--2~keV band by a factor of more than $\sim$160 relative to an archival 3$\sigma$ upper limit inferred from a serendipitous \textit{Chandra} pointing in 2011. The transient was also independently found in the optical by the Zwicky Transient Factory (ZTF), with the eRASS2 detection occurring only $\sim$20 days after the peak optical brightness, suggesting that the accretion disc formed promptly in this TDE. Continued X-ray monitoring over the following $\sim$400 days by eROSITA, \textit{NICER} XTI and \textit{Swift} XRT showed a net decline by a factor of $\sim$100, albeit with large amplitude X-ray variability where the system fades, and then rebrightens, in the 0.3--2~keV band by a factor $\sim$50 during an 80 day period. Contemporaneous \textit{Swift} UVOT observations during this extreme X-ray variability reveal a relatively smooth decline, which persists over $\sim$400 days post-optical peak. The peak observed optical luminosity (absolute $g$-band magnitude $\sim -16.8$ mag) from this transient makes J0744 the faintest optically-detected TDE observed to date. However, contrasting the known set of `faint and fast' TDEs, the optical emission from J0744 decays slowly (exponential decay timescale $\sim$120~days), making J0744 the first member of a potential new class of `faint and slow' TDEs.
\end{abstract}

\begin{keywords}
accretion, accretion discs -- galaxies: nuclei -- black hole physics -- transients: tidal disruption events --
\end{keywords}


\section{Introduction}
A star that passes within the tidal radius of a black hole (BH) will be torn apart by strong gravitational forces, in a so-called tidal disruption event (TDE; e.g.~\citealt{hills_possible_1975,rees_tidal_1988}). Ensuing this, the remaining bound portion of the stellar debris is elongated into long, thin streams, which evolve on highly eccentric orbits around the BH. Whilst aspects of accretion flow formation in TDEs are still uncertain (see \citealt{bonnerot_formation_2020} for a recent review), the disc formation process, and subsequent accretion of the stellar debris onto the BH, generates luminous electromagnetic flares that evolve over timescales of weeks-to-years, depending on the configuration of the TDE itself. These events can cause the accretion rate onto the black hole to transition from sub-Eddington, to super-Eddington, and back, over timescales of years, and thus may present excellent tools to probe both accretion physics, as well as the demographics of quiescent black holes.

The first strong TDE candidates \citep{bade_detection_1996,grupe_rx_1999,komossa_discovery_1999,komossa_giant_1999,greiner_rx_2000} were discovered by \textit{ROSAT} \citep{trumper_rosat_1982}, and were broadly consistent with the observational signatures of TDEs predicted in \citet{rees_tidal_1988}. Later discoveries of X-ray bright, non-jetted TDEs  \citep{maksym_tidal_2010,lin_discovery_2011,saxton_tidal_2012,maksym_tidal_2013,donato_tidal_2014,khabibullin_stellar_2014,lin_ultrasoft_2015,saxton_xmmsl1_2017,lin_likely_2017}, predominantly with \textit{XMM-Newton} \citep{jansen_xmm-newton_2001} and \textit{Chandra} \citep{weisskopf_chandra_2000}, built up a broad picture of these events as large amplitude, ultra-soft X-ray flares from galaxies that had shown no strong signatures of prior AGN activity. 

Several UV-bright TDEs were also identified in the 2000s \citep{gezari_ultraviolet_2006,gezari_uvoptical_2008,gezari_luminous_2009} with the \textit{Galaxy Evolution Explorer} (\textit{GALEX}; \citealt{martin_galaxy_2005}). Recent years have seen a significant increase in the reported detection rate in TDE candidates, with these having been primarily discovered through optical surveys such as ASAS-SN \citep{holoien_asassn-14ae_2014,leloudas_spectral_2019,wevers_evidence_2019,wevers_rapid_2021,hinkle_discovery_2020}, ATLAS \citep{wang_discovery_2022}, PanSTARRS \citep{holoien_ps18kh_2019}, PTF and iPTF \citep{arcavi_continuum_2014,blagorodnova_iptf16fnl_2017}, OGLE \citep{wyrzykowski_ogle16aaa_2017} and ZTF \citep{van_velzen_first_2019,van_velzen_seventeen_2021,nicholl_outflow_2020}. Curiously, the vast majority of optically-selected TDEs do not show any transient X-ray emission ($\sim$70\% of optically-selected TDEs in \citealt{van_velzen_seventeen_2021} were not X-ray `loud'). Possible mechanisms for explaining the dearth of X-rays from optically-selected TDEs include the optical emission being powered by stream-stream collisions, instead of accretion \citep{piran_disk_2015,shiokawa_general_2015}, or that the high-energy disc emission is being reprocessed to lower wavelengths, either through a quasi-spherical gaseous envelope (e.g.~\citealt{loeb_optical_1997}), or outflows (e.g.~disc winds, \citealt{lodato_multiband_2011,miller_disk_2015,metzger_bright_2016,dai_unified_2018}; collision-induced outflows, \citealt{lu_self-intersection_2020}). \citet{dai_unified_2018} suggest that the observed properties of a TDE may be highly dependent on the observer's viewing angle to the system, with X-ray bright TDEs being observed closer to pole-on inclinations, and optically bright TDEs being observed edge-on.

Launched in 2019, eROSITA \citep{predehl_erosita_2021}, on-board the Russian-German \textit{SRG} mission \citep{sunyaev_srg_2021}, has rapidly expanded the sample of X-ray selected TDE \textit{candidates} (e.g.~\citealt{brightman_luminous_2021,malyali_at_2021,sazonov_first_2021}, Malyali~et~al.~in~prep.). 
In this paper, we report on a set of multi-wavelength observations of the TDE candidate eRASSt J074426.5+29160 (herein J0744), associated to a previously inactive galaxy at $z=0.0396$. In section~\ref{sec:discovery}, we describe the initial discovery of J0744, before presenting an overview of X-ray observations of J0744 in section~\ref{sec:xray_observations}, and its photometric evolution in section~\ref{sec:photometric_evolution}. We then analyse the host galaxy properties of this TDE in section~\ref{sec:host_properties}, and discuss the unique aspects of this event, and implications for TDE science, in section~\ref{sec:discussion}. We conclude with a summary in section~\ref{sec:summary}.

We adopt a flat $\Lambda$CDM cosmology throughout this paper, with $H_0=67.7\, \mathrm{km}\, \mathrm{s}^{-1}\mathrm{Mpc}^{-1}$, $\Omega _{\mathrm{m}}=0.309$ \citep{planck_collaboration_planck_2016}; $z=0.0396$ therefore corresponds to a luminosity distance of 180.5~Mpc. All magnitudes will be reported in the AB system and corrected for Galactic foreground reddening using $A_{\mathrm{V}}=0.0902$~mag, obtained from \citep{schlafly_measuring_2011} and $R_{\mathrm{V}}=3.1$, and using the effective wavelength for each filter retrieved from the SVO Filter Profile Service\footnote{\url{http://svo2.cab.inta-csic.es/theory/fps/}}, unless otherwise stated. All dates/ times will be reported in universal time (UT) or MJD.

\begin{figure}
    \centering
    \includegraphics[scale=1.2]{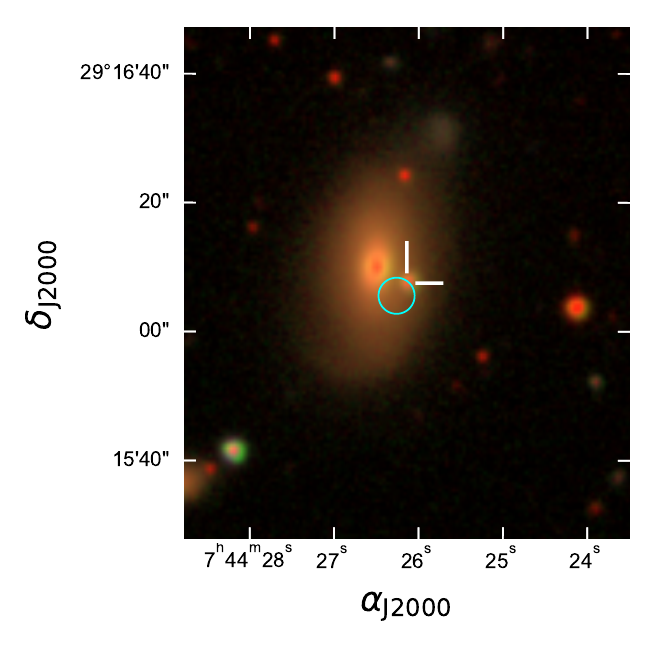}
    \caption{Legacy DR8 $g$, $r$, $z$ false colour image centred on the host galaxy of J0744. The white cross-hairs and cyan circle mark the ZTF position and eROSITA localisation respectively, where the radius of the circle is set to the 2$\sigma$ uncertainty on the eROSITA source position. The separation between the two galaxy centres is $\sim 5.4 ^{\prime \prime}$, corresponding to a projected distance of $\sim$4.3~kpc (section~\ref{sec:morphology}).}
    \label{fig:finder_chart}
\end{figure}
\section{Discovery}\label{sec:discovery}
J0744 was first identified during a systematic search for TDEs in the second eROSITA All-Sky Survey (eRASS2), where it was detected by eROSITA on 2020-10-21 as a new, bright (0.3--2~keV observed flux of $\sim 1\times 10^{-12}$~erg~cm$^{-2}$~s$^{-1}$), ultra-soft (photon index $\sim$4) X-ray point source, which had not been detected 6 months previously in eRASS1 (section~\ref{sec:erosita_observations}). Using the \textit{eROSITA Science Analysis Software} pipeline (eSASS; \citealt{brunner_erosita_2022}), the source was localised to (RA$_\mathrm{J2000}$, Dec$_\mathrm{J2000})$=(07h44m26.3s,+29$^{\circ}$16$^\prime$04.8$^{\prime\prime}$), 
with a 1$\sigma$ positional uncertainty of 1.7$^{\prime \prime}$ (68\% confidence), and thus consistent with a pair of galaxies at $z=0.0396$ (Fig.~\ref{fig:finder_chart}). Based only on this localisation provided by eROSITA, then the optical counterpart was initially unclear. 
However, on 2020-10-02, the ALeRCE alert broker \citep{forster_automatic_2021} discovered the optical transient, AT~2020uwq/ ZTF20acgrymn\footnote{\url{https://www.wis-tns.org/object/2020uwq}, reported to the TNS on 2020-10-04.} within the Zwicky Transient Facility survey data (ZTF; \citealt{bellm_zwicky_2019,graham_zwicky_2019}), which is associated with the centre of the smaller galaxy in Fig.~\ref{fig:finder_chart} (herein SDSS~J074426.12+291607.4- section~\ref{sec:host_properties}). As it is highly unlikely that there also exists an ultra-soft X-ray transient (section~\ref{sec:xray_observations}) evolving contemporaneously in the more massive galaxy seen in Fig.~\ref{fig:finder_chart}, we consider the transient X-ray emission initially detected by eROSITA to also originate from SDSS~J074426.12+291607.4 throughout this work.

\section{X-ray observations}\label{sec:xray_observations}
\subsection{Archival}\label{sec:archival_xray}
The position of J0744 was serendipitously observed by \textit{Chandra} during a 10.8~ks exposure on 2011-02-07 with ACIS-S. No X-ray point source was detected at the position of the host galaxy in this observation. Using the \texttt{ciao} \citep{fruscione_ciao_2006} tool \texttt{srcflux}, the 3$\sigma$ upper limit on the 0.3--2~keV flux of this source is $2.6 \times 10^{-15}$~erg~cm$^{-2}$~s$^{-1}$, assuming the same spectral model as the best-fit spectral model to eROSITA spectra described in section~\ref{sec:xray_spectral_fitting}. The nucleus of the more massive galaxy was detected as a faint point source in the \textit{Chandra} Source Catalog 2(CSC2) 
located at (RA$_\mathrm{J2000}$, Dec$_\mathrm{J2000})$=(07h44m26.5s,+29$^{\circ}$16$^\prime$10.2$^{\prime\prime}$), with a positional uncertainty of 1.3$^{\prime \prime}$ (95\% confidence) and with 0.5-7~keV flux of $(1.57 \pm 0.07)\times 10^{-14}$~erg~cm$^{-2}$~s$^{-1}$. However, the X-ray emission from this source is hard, with no source detection reported in the 0.2--0.5~keV or 0.5--1.2~keV bands in CSC2. Assuming an AGN-like spectral model with $\Gamma =2$ (e.g.~ \citealt{nandra_ginga_1994}), and absorption set to the Galactic neutral hydrogen equivalent column density along the line of sight, the 3$\sigma$ UL on the source's 0.3--2~keV flux is $6 \times 10^{-15}$~erg~cm$^{-2}$~s$^{-1}$. In addition to the \textit{Chandra} observations, no X-ray source was detected within 60$^{\prime\prime}$ of J0744 during \textit{ROSAT} observations during the 1990s, nor during \textit{Neil Gehrels Swift} XRT \citep{gehrels_swift_2004,burrows_swift_2005} observations in 2014. Both of these provide weaker constraints on the archival X-ray flux than the 2011 \textit{Chandra} observation.

\begin{table}
\centering
\caption{Log of X-ray observations of J0744. For eROSITA, the mid-date of the coverage in each eRASS is given.} 
\label{tab:x_ray_observation_log}
\begin{tabular}{cccc}
\hline
MJD & Instrument & ObsID & Exposure [s] \\
\hline
58957.587 & \textit{SRG}/eROSITA & eRASS1 & 147 \\
59143.580 & \textit{SRG}/eROSITA & eRASS2 & 127 \\
59164.456 & \textit{Swift}/XRT & 00013855001 & 790 \\
59176.821 & \textit{Swift}/XRT & 00013855003 & 1173 \\
59179.749 & \textit{Swift}/XRT & 00013855004 & 2084 \\
59181.669 & \textit{NICER}/XTI & 3201920101 & 8117 \\
59182.057 & \textit{NICER}/XTI & 3201920102 & 4674 \\
59183.111 & \textit{NICER}/XTI & 3201920103 & 533 \\
59185.086 & \textit{Swift}/XRT & 00013855005 & 2091 \\
59185.596 & \textit{NICER}/XTI & 3201920104 & 134 \\
59202.861 & \textit{Swift}/XRT & 00013855006 & 1362 \\
59212.289 & \textit{Swift}/XRT & 00013855008 & 1647 \\
59220.631 & \textit{Swift}/XRT & 00013855009 & 1063 \\
59233.053 & \textit{Swift}/XRT & 00013855011 & 1279 \\
59239.732 & \textit{Swift}/XRT & 00013855012 & 998 \\
59259.083 & \textit{Swift}/XRT & 00013855013 & 800 \\
59286.060 & \textit{Swift}/XRT & 00013855014 & 888 \\
59314.271 & \textit{Swift}/XRT & 00013855015 & 837 \\
59323.879 & \textit{SRG}/eROSITA & eRASS3 & 198 \\
59342.744 & \textit{Swift}/XRT & 00013855016 & 2061 \\
59509.205 & \textit{SRG}/eROSITA & eRASS4 & 177 \\
59559.675 & \textit{Swift}/XRT & 00013855017 & 2001 \\
\end{tabular}
\end{table}

\subsection{eROSITA}\label{sec:erosita_observations}
\begin{figure*}
    \centering
    \includegraphics{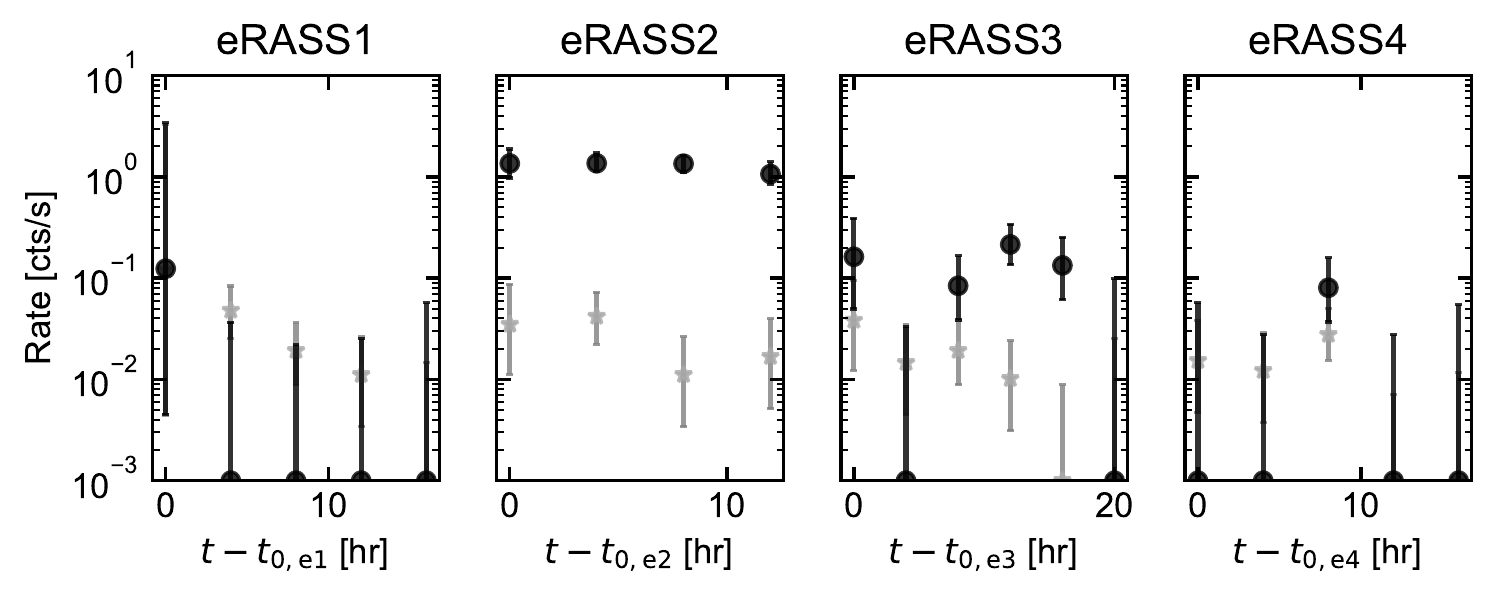}
    \caption{0.2--2.3keV band eROSITA lightcurves of J0744, with each sub plot showing the X-ray evolution within a given eRASS. Within an eRASS scan, each successive visit is $\sim$4 hours after the previous visit, whilst each successive eRASS is $\sim$6 months after the previous eRASS. The black markers represent the count rate measured within the source aperture (not-background subtracted), whilst grey markers denote the background count rate measured within the background aperture (re-scaled to the size of the source aperture). $t-t_{0,i}$ represents the time taken since the start of observations of J0744 within eRASS $i$. }
    \label{fig:intra_erass_lightcurve}
\end{figure*}

J0744 was detected by the eROSITA source detection pipeline as a new ultra-soft X-ray source on 2020-10-21 during eRASS2, with a 0.3--2~keV count rate of $1.05^{+0.14}_{-0.13}$ ct~s$^{-1}$. No source was detected by eROSITA when it scanned the position of J0744 on 2020-04-18 in eRASS1, with a 3$\sigma$ upper limit on the 0.3--2~keV count rate 0.04~ct~s$^{-1}$. J0744 was also later observed by eROSITA during eRASS3 and eRASS4 between 2021-04-19 and 2021-04-20, and 2021-10-21 and 2021-10-22, respectively. 
\begin{table}
\centering
\caption{eROSITA count data of J0744 in the 0.3--2~keV band. $N_{\mathrm{src}}$ is the total number of counts detected in the source aperture, exposure is the non-vignetting corrected exposure of the source within a given eRASS, $N_{\mathrm{bkg}}$ is the total number of counts in the background aperture, scaled to the area of the source aperture. $CR_{\mathrm{src}}$ is the source count rate posterior median, with the superscript and subscript values denoting the bounds of the 1$\sigma$ credible region. Upper limits are provided for observations with a detection significance below 3$\sigma$.}
\label{tab:erosita_count_rates}
\begin{tabular}{cccccc}
\hline
MJD & ObsID & Exposure [s] & $N_{\mathrm{src}}$ & $N_{\mathrm{bkg}}$ & $CR_{\mathrm{src}}$ [ct s$^{-1}$]\\
\hline
58957.587 & eRASS1 & 147 & 0 & 1.7 & $<$0.04 \\
59143.580 & eRASS2 & 127 & 64 & 2.0 & 1.05$^{+0.14}_{-0.13}$ \\
59323.879 & eRASS3 & 198 & 9 & 2.9 & 0.05$^{+0.04}_{-0.04}$ \\
59509.205 & eRASS4 & 177 & 2 & 1.8 & $<$0.06
\end{tabular}
\end{table}
For each observation of J0744 within each eRASS, spectra and lightcurves were extracted using the \texttt{SRCTOOL} task provided by eSASS (version eSASS\_users211214). To extract source spectra, a circular aperture of radius 60$^{\prime \prime}$\footnote{Whilst this aperture will be contaminated by flux from the larger galaxy, its 0.3--2~keV flux inferred from \textit{Chandra} (section~\ref{sec:archival_xray}) is well below the sensitivity of eROSITA and thus not expected to contribute significantly to the counts measured for the TDE.}, centred on the eRASS2 position, was used, whilst background spectra and lightcurves were extracted from a source-free annulus with inner and outer radii of 140$^{\prime \prime}$ and 240$^{\prime \prime}$, respectively. No significant variability is seen in the 0.2--2.3~keV band \textit{within} the four observations of J0744 during eRASS2 (Fig.~\ref{fig:intra_erass_lightcurve}). 

The eROSITA 0.3--2~keV count rates were converted to 0.3--2~keV observed fluxes using an energy conversion factor (ECF) of $9.4\times 10^{-13}$~erg~cm$^{-2}$, and $1.2 \times 10^{-12}$~erg~cm$^{-2}$ for the fluxes corrected for Galactic absorption, which were computed using the best fitting spectral model to the eRASS2 observation (section~\ref{sec:xray_spectral_fitting}). The 0.3--2~keV flux of this source in the eRASS2 observation is $9.8^{+1.3}_{-1.2} \times 10^{-13}$~erg~cm$^{-2}$~s$^{-1}$, thus the source had brightened by a factor of at least $\sim$160 relative to the 3$\sigma$ upper limit from the archival Chandra observation in 2011.

\begin{table}
\centering
\caption{Inferred fluxes in the 0.3--2~keV band for J0744, with $F_{\rm 0.3-2keV, unabs}$ and $F_{\rm 0.3-2keV, obs}$ being corrected and not corrected for Galactic absorption, respectively. Upper limits are provided for sources with a detection significance below 3$\sigma$.}
\label{tab:flux_table}
\begin{tabular}{ccccc}
\hline
MJD & Instrument & ObsID & $F_{\rm 0.3-2keV, obs}$  & $F_{\rm 0.3-2keV, unabs}$ \\
 & & & [$10^{-13}$\,erg cm$^{-2}$ s$^{-1}$ & [$10^{-13}$\,erg cm$^{-2}$ s$^{-1}$] \\
\hline
58957.587 & eROSITA & eRASS1 & $<$0.4 & $<$0.5 \\
59143.580 & eROSITA & eRASS2 & $9.8 ^{+1.3}_{-1.2}$ & $12.8 ^{+1.7}_{-1.6}$ \\
59164.456 & XRT & 00013855001 & $<$1.9 & $<$2.3 \\
59176.821 & XRT & 00013855003 & $<$2.1 & $<$2.6 \\
59179.749 & XRT & 00013855004 & $<$2.3 & $<$2.7 \\
59181.669 & XTI & 3201920101 & $<$4.1 & $<$4.9 \\
59182.057 & XTI & 3201920102 & $<$4.6 & $<$5.6 \\
59183.111 & XTI & 3201920103 & $<$8.5 & $<$10.3 \\
59185.086 & XRT & 00013855005 & $<$1.0 & $<$1.2 \\
59185.596 & XTI & 3201920104 & $<$7.0 & $<$8.5 \\
59202.861 & XRT & 00013855006 & $1.1 ^{+0.6}_{-0.4}$ & $1.3 ^{+0.7}_{-0.5}$ \\
59212.289 & XRT & 00013855008 & $4.1 ^{+0.8}_{-0.8}$ & $4.9 ^{+1.0}_{-1.0}$ \\
59220.631 & XRT & 00013855009 & $10.4 ^{+2.0}_{-2.0}$ & $12.4 ^{+2.4}_{-2.4}$ \\
59233.053 & XRT & 00013855011 & $<$1.5 & $<$1.7 \\
59239.732 & XRT & 00013855012 & $<$2.3 & $<$2.7 \\
59259.083 & XRT & 00013855013 & $<$3.4 & $<$4.1 \\
59286.060 & XRT & 00013855014 & $<$5.5 & $<$6.5 \\
59314.271 & XRT & 00013855015 & $<$3.0 & $<$3.6 \\
59323.879 & eROSITA & eRASS3 & $0.5 ^{+0.4}_{-0.4}$ & $0.6 ^{+0.5}_{-0.5}$ \\
59342.744 & XRT & 00013855016 & $0.9 ^{+0.6}_{-0.4}$ & $1.1 ^{+0.7}_{-0.5}$ \\
59509.205 & eROSITA & eRASS4 & $<$0.6 & $<$0.8 \\
59559.675 & XRT & 00013855017 & $<$1.7 & $<$2.0 \\
\end{tabular}
\end{table}

\subsection{\textit{Swift} XRT}
Following the eROSITA detection, J0744 was further monitored by the XRT and UVOT \citep{roming_swift_2005} instruments on board the \textit{Neil Gehrels Swift} observatory (Table~\ref{tab:x_ray_observation_log}). Observations taken with XRT were performed in photon counting mode. The XRT lightcurves were generated and downloaded from the UK Swift Science Data Centre website \citep{evans_online_2007,evans_methods_2009}, and were analysed using version 6.29 of the HEASoft analysis software, with CALDB v20210915. The source count rates and upper limits provided by this tool were inferred using the \citet{kraft_determination_1991} method, and have already been corrected for pile-up, CCD dead zones, and the loss of photons outside of the source extraction regions \citep{evans_methods_2009}. For the two XRT observations where more than 20 counts were detected within the source extraction region, \texttt{xrtproducts} was also used to generate source and background spectra. To do this, source counts were extracted from a circular aperture of 47$^{\prime \prime}$ radius, with background counts extracted from an annulus with inner and outer radii 70$^{\prime \prime}$ and 250$^{\prime \prime}$, respectively. The 0.3--1~keV XRT count rates were then converted to observed 0.3--2~keV fluxes using an ECF of $2.1\times 10^{-11}$~erg~cm$^{-2}$, and $2.1\times 10^{-11}$~erg~cm$^{-2}$ for the 0.3--2~keV fluxes corrected for Galactic absorption, which was again computed using the best fitting spectral model for the XRT observation with OBSID 00013855009 (section~\ref{sec:xray_spectral_fitting}). 
\begin{table}
\centering
\caption{\textit{Swift}/XRT count data of J0744 in the 0.3--1~keV band. The column definitions are the same as for table~\ref{tab:erosita_count_rates}, upper limits are provided for sources with a detection significance below 3$\sigma$.}
\label{tab:xrt_count_rates}
\begin{tabular}{cccccc}
\hline
MJD & ObsID & Exposure [s] & $N_{\mathrm{src}}$ & $N_{\mathrm{bkg}}$ & $CR_{\mathrm{src}}$ [ct s$^{-1}$]\\
\hline
59164.456 & 00013855001 & 790 & 0 & 0.0 & $<$0.009 \\
59176.821 & 00013855003 & 1173 & 0 & 0.0 & $<$0.010 \\
59179.749 & 00013855004 & 2084 & 1 & 0.1 & $<$0.011 \\
59185.086 & 00013855005 & 2091 & 0 & 0.1 & $<$0.005 \\
59202.861 & 00013855006 & 1361 & 5 & 0.1 & 0.005$^{+0.003}_{-0.002}$ \\
59212.289 & 00013855008 & 1647 & 24 & 0.3 & 0.019$^{+0.004}_{-0.004}$ \\
59220.631 & 00013855009 & 1063 & 28 & 0.2 & 0.049$^{+0.009}_{-0.009}$ \\
59233.053 & 00013855011 & 1279 & 0 & 0.1 & $<$0.007 \\
59239.732 & 00013855012 & 998 & 1 & 0.2 & $<$0.011 \\
59259.083 & 00013855013 & 800 & 0 & 0.1 & $<$0.016 \\
59286.060 & 00013855014 & 888 & 0 & 0.0 & $<$0.026 \\
59314.271 & 00013855015 & 837 & 1 & 0.1 & $<$0.014 \\
59342.744 & 00013855016 & 2061 & 4 & 0.2 & 0.004$^{+0.003}_{-0.002}$ \\
59559.675 & 00013855017 & 2001 & 0 & 0.1 & $<$0.008 \\
\end{tabular}
\end{table}

\subsection{\textit{NICER} XTI}\label{sec:nicer}
J0744 was also monitored over a four day window between 2020-11-28 and 2020-12-02 (Table~\ref{tab:x_ray_observation_log}) by the X-ray Timing Instrument (XTI) on-board the \textit{Neutron Star Interior Composition Explorer} (\textit{NICER}; \citealt{den_herder_neutron_2016}). The observational data was downloaded from the \textit{NICER} data archive available through HEASARC, with which we then ran the \texttt{nicerl2} task on to generate a set of cleaned event files. For each OBSID, the \texttt{nibackgen3C50} tool \citep{remillard_empirical_2022} was then used to generate source and (empirically generated) background spectra from the cleaned event files. From these, we inferred total and background count rates in the 0.3-1~keV band, with this band chosen to reduce background contamination of the source count rate due to incomplete background modelling, and because of the source's ultra-soft X-ray spectrum (section~\ref{sec:xray_spectral_fitting}). J0744 is background-dominated in these observations (Table~\ref{tab:nicer_x_ray_rates}), with 3$\sigma$ upper limits on the count rates conservatively inferred via $CR_{\mathrm{tot}}+3\sigma$, where $\sigma$ is the error on the total count rate. The 0.3-1~keV band count rates were then converted to 0.3--2~keV fluxes (Table~\ref{tab:flux_table}) assuming the best-fitting spectral model to the eRASS2 observation.

\begin{table}
\centering
\caption{\textit{NICER}/XTI count data of J0744 in the 0.3-1~keV band. The $CR_{\mathrm{tot}}$ and $CR_{\mathrm{bkg}}$ are the total and background estimated count rates inferred via the \texttt{nibackgen3C50} tool. We conservatively estimate 3$\sigma$ upper limits on the count rate using $CR_{\mathrm{tot}}+3\sigma$, where $\sigma$ is the error on the total count rate}.
\label{tab:nicer_x_ray_rates}
\begin{tabular}{ccccc}
\hline
MJD & ObsID &Exposure [s] &  $CR_{\mathrm{tot}}$ [ct s$^{-1}$] &  $CR_{\mathrm{bkg}}$ [ct s$^{-1}$]\\
\hline
59181.669 & 3201920101 & 8117 & $0.25 \pm 0.07 $ & 0.21 \\
59182.057 & 3201920102 & 4674 & $0.27 \pm 0.09 $ & 0.24 \\
59183.111 & 3201920103 & 533 & $0.47 \pm 0.17 $ & 0.34 \\
59185.596 & 3201920104 & 134 & $0.22 \pm 0.20 $ & 0.13 \\
\hline
\end{tabular}
\end{table}

\subsection{Spectral fitting}\label{sec:xray_spectral_fitting}
The extracted X-ray spectra were analysed using the Bayesian X-ray Analysis software (BXA; \citealt{buchner_x-ray_2014}), which connects the fitting environment XSPEC \citep{arnaud_xspec_1996} with the nested sampling algorithm UltraNest\footnote{\url{https://johannesbuchner.github.io/UltraNest/}} \citep{buchner_ultranest_2021}; spectra were fitted unbinned using the C-statistic \citep{cash_generation_1976}. 

The eROSITA and XRT backgrounds were first modelled empirically using the principal component analysis (PCA) technique described in \citet{simmonds_xz_2018} (see also \citealt{liu_erosita_2022}). The extracted source spectra, which contain a contribution from both the source and background emission, were then jointly fitted using the source and background models. However, due to the combination of the relatively faint X-ray flux of the source, and the short eROSITA and XRT exposures, only the eRASS2 and Swift observations 00013855008 and 00013855009 had more than 20 counts within the source extraction regions (Table~\ref{tab:erosita_count_rates} and \ref{tab:xrt_count_rates}) and were fitted with BXA. 

The eRASS2 X-ray spectrum of J0744 is ultra-soft, with nearly all counts in the source aperture being detected below 2~keV. Five different models were fitted to this spectrum over the 0.2-8~keV range: (i) a redshifted blackbody (\texttt{tbabs*zbbody}), (ii) a redshifted power-law (\texttt{tbabs*zpwrlaw}), iii) an absorbed redshifted blackbody (\texttt{tbabs*zabsbb}), iv) an absorbed redshifted powerlaw (\texttt{tbabs*zabspwrlaw}) and v) redshifted thermal bremsstrahlung\footnote{The high-energy tail of bremmstrahlung emission may produce an ultra-soft X-ray spectrum, broader than that of a single temperature blackbody. This has previously been used for fitting the X-ray spectra of TDEs (e.g. \citealt{saxton_tidal_2012,wevers_rapid_2021}), although the physical origin of the bremmstrahlung emission is still unclear.} (\texttt{tbabs*zbremss}). 
For each fitting, the equivalent Galactic neutral hydrogen column density, $N_{\mathrm{H}}$, for the \texttt{tbabs} component, was set to its value in the HI4PI survey of $3.52 \times 10^{20}$ cm$^{-2}$ \citep{hi4pi_collaboration_hi4pi_2016}, and the redshift was fixed to that of the host galaxy. The same procedure was adopted for the fitting of the XRT spectra, except that these were fitted over the 0.3-8~keV range. The relative values of the Bayesian evidence, computed by UltraNest, is used to choose the best fitting model to each spectrum (see \citealt{buchner_x-ray_2014}), with these listed along with other parameter estimates obtained from BXA in Table~\ref{tab:x_ray_model_fits}.

\begin{figure}
    \centering
    \includegraphics{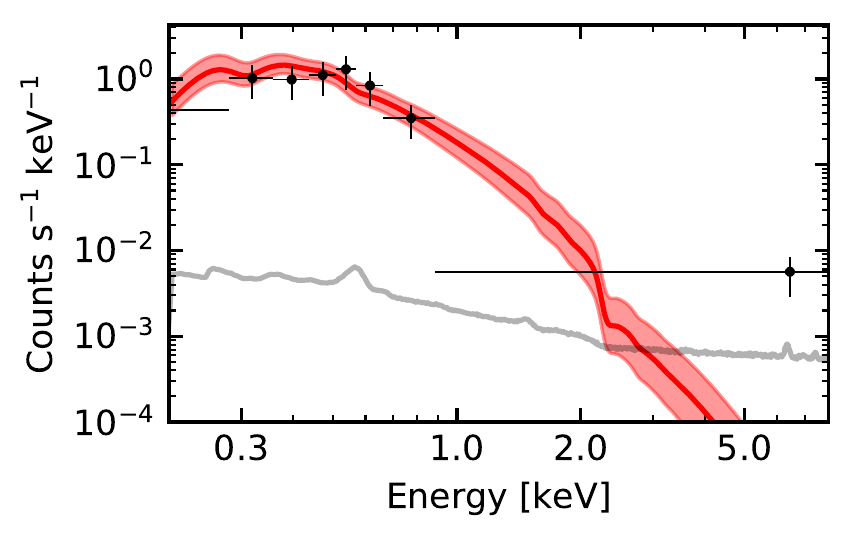}
    \caption{BXA fit of the \texttt{tbabs*zpwrlaw} model to the eRASS2 source spectrum of J0744, where the darker (lighter) red filled bands enclose 68\% (98\%) of the posterior. The X-ray spectrum is ultra-soft, with $\Gamma=4.0^{+0.2}_{-0.3}$. Black markers denote the source count rates, with counts being binned to 10 counts per bin for plotting purposes only; the BXA fit was performed on the unbinned spectrum. The solid grey line represents the median of the best-fitting background model inferred from the BXA fit.}
    \label{fig:convolved_posterior_ztbabs_zpowerlaw}
\end{figure}

The best fitting spectral model to the eRASS2 spectrum is \texttt{tbabs*zpwrlaw}, with the inferred photon index, $\Gamma=4.0 ^{+0.2}_{-0.3}$, and the convolved source model is plotted in Fig.~\ref{fig:convolved_posterior_ztbabs_zpowerlaw}. \texttt{tbabs*zpwrlaw} is also the preferred model for the XRT observation obtained $\sim$70 days after the eRASS2 observation (OBSID 00013855008), with the photon index, $3.8 ^{+0.4}_{-0.4}$, consistent with the eRASS2 observation. Whilst the preferred model for the 00013855009 spectrum is \texttt{tbabs*zbbody}, with $kT=0.14^{+0.02}_{-0.01}$~keV, overall there is no evidence for any major X-ray spectral changes between the eRASS2, 00013855008 and 00013855009 spectra, when looking at the parameter estimates for a given source model (e.g. $kT$ are consistent with each other for \texttt{tbabs*zbbody} across the three spectra in Table~\ref{tab:x_ray_model_fits}). 
\begin{table*}
\centering
\caption{X-ray spectral fit results inferred using BXA, with the abbreviated model names representing the \texttt{tbabs*zbbody}, \texttt{tbabs*zpwrlaw}, \texttt{tbabs*zabsbb},  \texttt{tbabs*zabspwrlaw} and \texttt{tbabs*zbremss} models respectively (defined in section~\ref{sec:xray_spectral_fitting}). The $kT$ columns are in units of keV, and $N_{\mathrm{H}}$ in $10^{22}$~cm$^{-2}$. $\Delta \mathcal{Z}$ is provided for each fitted source model, and is defined as the log of the evidence ratio between that model, and the model with the highest evidence in the row, $\Delta \mathcal{Z} = 0$ therefore indicates that a model is the `best fitting' to the observed X-ray spectrum. The lower the value of $\Delta \mathcal{Z}$, the more strongly this model is disfavoured relative to the best fitting model.}
\label{tab:x_ray_model_fits}
\begin{tabular}{c|cccccccccccccc}
\hline
MJD & OBSID & \multicolumn{2}{c}{\texttt{zbb}} & \multicolumn{2}{c}{\texttt{zpo}} & \multicolumn{3}{c}{\texttt{zabsbb}} & \multicolumn{3}{c}{\texttt{zabspo}} & \multicolumn{3}{c}{\texttt{zbrems}} \\
\hline
 & & $\Delta \mathcal{Z}$ & $kT$ & $\Delta \mathcal{Z}$ & $\Gamma$ & $\Delta \mathcal{Z}$ & $N_{\mathrm{H}}$ & $kT$ & $\Delta \mathcal{Z}$ & $N_{\mathrm{H}}$& $\Gamma$ & $\Delta \mathcal{Z}$ & $kT_{\mathrm{p}}$  \\
\hline
59143.580 & eRASS2 & -4.3 & $0.10 ^{0.01}_{0.01}$ & 0.0 & $4.0 ^{0.2}_{0.3}$ & -8.6 & $0.01 ^{0.01}_{0.00}$ & $0.10 ^{0.01}_{0.01}$ & -1.7 & $0.02 ^{0.02}_{0.01}$ & $4.4 ^{0.4}_{0.3}$ & -1.3 & $0.22 ^{0.03}_{0.02}$ \\
59212.289 & 00013855008 & -7.6 & $0.15 ^{0.02}_{0.02}$ & 0.0 & $3.8 ^{0.4}_{0.4}$ & -9.8 & $0.02 ^{0.03}_{0.01}$ & $0.14 ^{0.02}_{0.02}$ & -1.4 & $0.04 ^{0.08}_{0.03}$ & $4.2 ^{0.7}_{0.5}$ & -2.3 & $0.41 ^{0.11}_{0.08}$ \\
59220.631 & 00013855009 & 0.0 & $0.14 ^{0.02}_{0.01}$ & -3.9 & $3.7 ^{0.4}_{0.4}$ & -0.8 & $0.08 ^{0.17}_{0.06}$ & $0.13 ^{0.02}_{0.02}$ & -1.2 & $0.42 ^{0.16}_{0.19}$ & $6.7 ^{1.0}_{1.4}$ & -1.0 & $0.35 ^{0.08}_{0.06}$ \\
\hline
\end{tabular}
\end{table*}

\subsection{Long term X-ray variability}\label{sec:xray_long_term_variability}
\begin{figure*}
    \centering
    \includegraphics[scale=0.75]{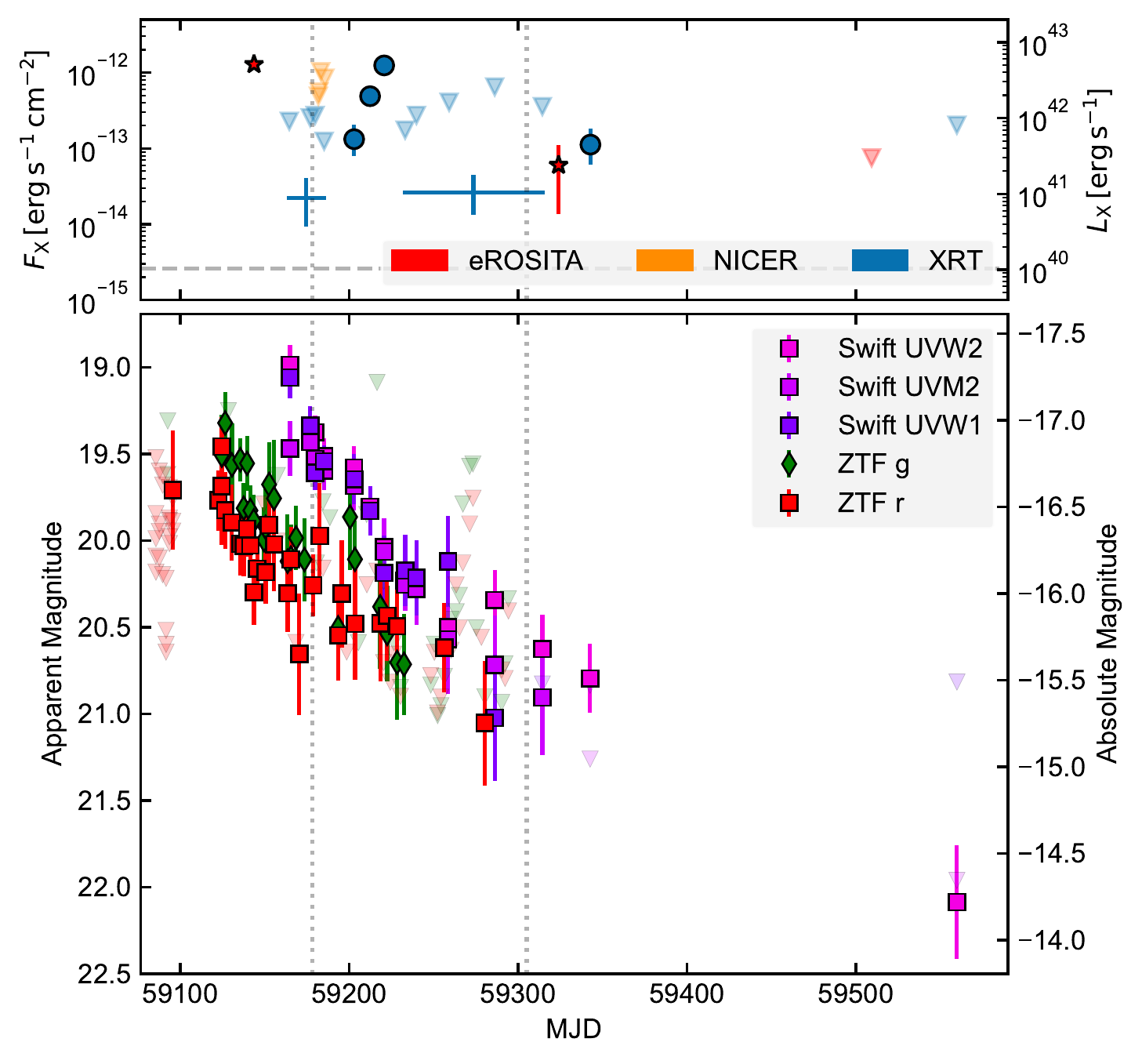}
    \caption{Multi-wavelength evolution of J0744. The top panel shows the X-ray lightcurve evolution in the 0.3--2~keV band, with \textit{eROSITA} (red), \textit{NICER} (orange), and \textit{Swift} XRT (blue) data points. The grey horizontal dashed line in this panel marks the 3$\sigma$ UL on the 0.3--2~keV source flux inferred from a 2011 Chandra pointing, when the source was previously non-detected; the eRASS2 detection represents a brightening by a factor of $\sim$160 relative to this limit. 3$\sigma$ upper limits on the source flux during eROSITA and XRT observations are plotted using triangles.  The blue crosses in the top panel represent the inferred fluxes from the stacked XRT observations (ObsIDs 00013855101 and 00013855102). The bottom panel shows the ZTF and \textit{Swift UVOT} lightcurve. Vertical grey dotted lines denote the times of the optical spectra being taken (section~\ref{sec:optical_spec}).}
\label{fig:multiwavelength_evolution}
\end{figure*}

Both the XRT and XTI observations improve the sampling of the X-ray lightcurve, revealing extreme, large amplitude X-ray variability in between the eROSITA observations of J0744 (Fig.~\ref{fig:multiwavelength_evolution}). To further constrain the evolution of the soft X-ray flux during the XRT monitoring campaign, we stacked the four XRT observations taken between 2020-11-11 and 2020-12-03 (XRT observation IDs 00013855001, 00013855003, 00013855004 and 00013855005), and 2021-01-18 and 2021-04-10 (XRT observation IDs 00013855011, 00013855012, 00013855013, 00013855014 and 00013855015), summing to net exposures of $6.1$~ks and $4.8$~ks, respectively. We assign these two stacked observations the IDs 00013855101 and 00013855102. Fitting the spectra extracted from these two observations with a \texttt{tbabs*zbbody} model with $kT$ fixed to 0.14~keV, then the inferred 0.3--2~keV X-ray flux in these stacked observations are $2.2^{+1.6}_{-1.1} \times 10^{-14}$~erg~cm$^{-2}$~s$^{-1}$ and $2.6^{+1.9}_{-1.2} \times 10^{-14}$~erg~cm$^{-2}$~s$^{-1}$, respectively.

 This implies that the system fades, and rebrightens, in the 0.3--2~keV band by a factor $\sim$50 over an 80 day period (Fig.~\ref{fig:multiwavelength_evolution}). Over the three XRT observations preceding the brightest epoch observed by XRT on MJD 59220, the system brightens by a factor of $\sim$10 over an 18 day period. This peak brightness is subsequently followed by another drop-off in the X-ray flux and a later rebrightening episode (Fig.~\ref{fig:multiwavelength_evolution}). At late times (one year after the eRASS2 observation), the system is non-detected with eROSITA during eRASS4, and also with XRT.

\section{Photometric evolution}\label{sec:photometric_evolution}
\subsection{Optical}\label{sec:ZTF}
We downloaded the publicly available ZTF light curve data of J0744 using the ZTF forced photometry service \citep{masci_zwicky_2019}, with the ZTF light curves provided through this interface having been already reference image subtracted. The raw forced photometry lightcurves were then calibrated using the method developed by Miller et al. (in prep.) for the ZTF Bright Transient Survey\footnote{\url{https://github.com/BrightTransientSurvey/ztf_forced_phot}}, which involved flagging and removing unreliable measurements, subtracting the pre-outburst baseline fluxes, and re-scaling of flux uncertainties. The ZTF photometry is presented in Fig.~\ref{fig:multiwavelength_evolution} and Table~\ref{tab:full_photometry}.

\begin{figure*}
    \centering
    \includegraphics[scale=0.8]{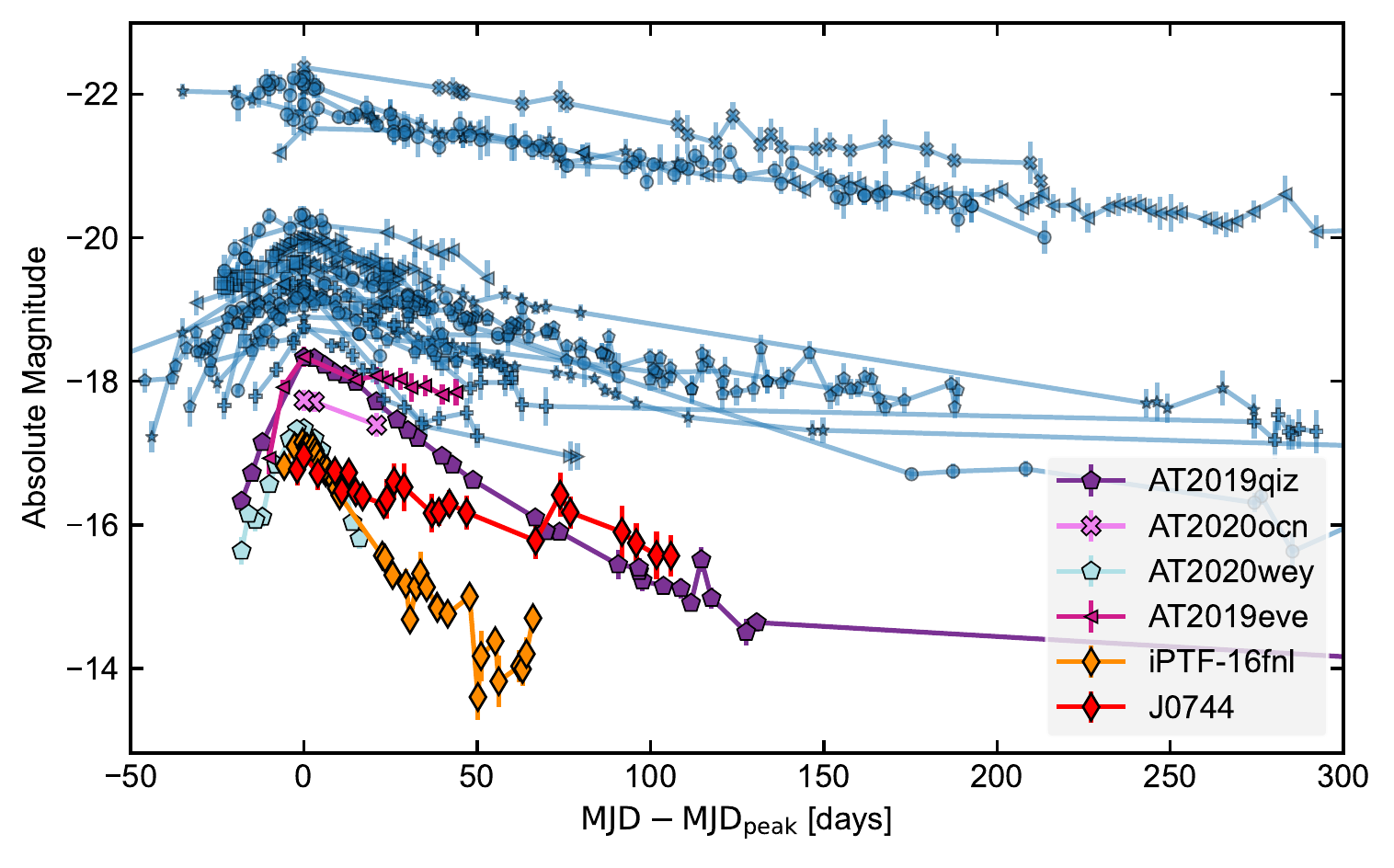}
    \caption{Comparison of the ZTF $g$-band lightcurve of J0744 with other optically bright TDE candidates (blue markers) reported in the ZTF survey (\citealt{van_velzen_seventeen_2021,hammerstein_final_2022}), and also with the faint and fast TDE iPTF-16fnl \citep{blagorodnova_iptf16fnl_2017}. The peak observed $g$-band brightness of J0744 is the faintest of all known TDEs identified to date. The line connecting each photometric datapoint for each TDE is only included to help guide the eye. }
    \label{fig:tde_lc_comparison}
\end{figure*}

\begin{figure}
    \centering
    \includegraphics[scale=0.8]{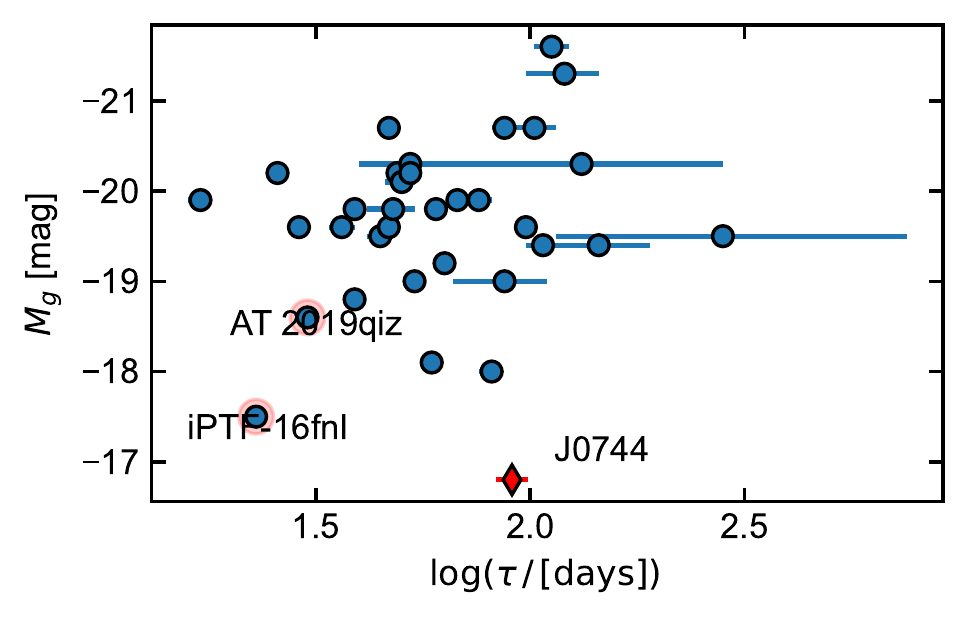}
    \caption{Peak inferred absolute $g$-band magnitude, $M_g$, against the exponential decay timescale, $\tau$, for the population of optically-bright TDEs (blue circle markers) presented in Tables 1 and 2 of \citet{van_velzen_optical-ultraviolet_2020}. J0744 (red diamond) is extremely faint ($M_g \sim -16.8$) relative to the bulk of this TDE population. The `faint and fast' TDEs iPTF-16fnl and AT~2019qiz are highlighted here with a red outline.}
    \label{fig:J0744_tde_comparison_absmag_vs_decay_timescale}
\end{figure}

\begin{figure}
    \centering
    \includegraphics[scale=0.8]{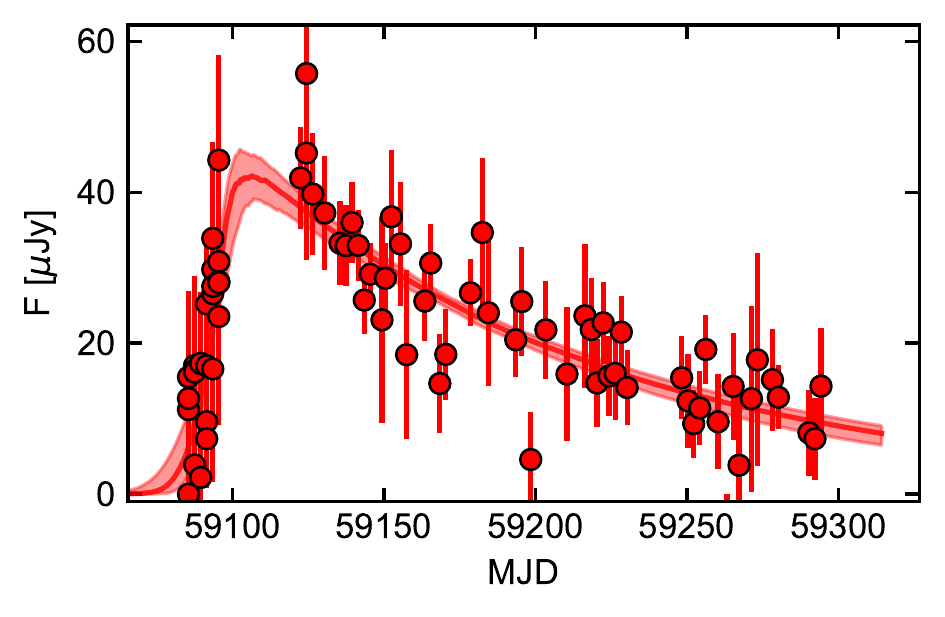}
    \caption{Fit to the calibrated ZTF $r$-band forced photometry. The inferred rise timescale for the $r$-band lightcurve is $\sim 10$~days, whilst the exponential decay timescale, $\tau$, is $\sim 120$~days.}
    \label{fig:ztf_r_band_exponential_decay_fit}
\end{figure}

J0744 was observed at peak brightness in the \textit{g}-band of $19.6 \pm 0.2$~mag on 2020-10-04, $\sim$17 days before the eROSITA eRASS2 observation. The optical emission then declined in flux by $\sim$0.5 mag in the following $\sim$20 days. The peak observed brightness is faint with respect to both its apparent and absolute magnitude, with the former likely being partially explainable for why the transient was missed in the larger ZTF TDE sample (i.e. it did not show a large amplitude optical flare similar to other TDEs). In terms of its peak absolute $g$-band magnitude ($\sim -16.8$ mag), it is also the optically faintest TDE reported to date (Fig.~\ref{fig:tde_lc_comparison}), out of the TDEs which have had a detection of their transient optical emission. Furthermore, whilst the other detected TDEs to-date with similar peak luminosities have been characterised as `faint and fast' (e.g. iPTF-16fnl, \citealt{blagorodnova_iptf16fnl_2017}; AT~2019qiz, \citealt{nicholl_outflow_2020}), the optical lightcurve of J0744 evolves over much slower timescales in comparison (Fig.~\ref{fig:J0744_tde_comparison_absmag_vs_decay_timescale}). Fitting the $r$-band lightcurve with a half-Gaussian function for the rise, and an exponential function for the decay (i.e.~following the approach described in section~3.1. of \citealt{malyali_at_2021}), then the inferred rise timescale for the $r$-band lightcurve is $10 ^{+6}_{-3}$ days, and the exponential decay timescale, $\tau$,  is $120 ^{+15}_{-12}$~days (Fig.~\ref{fig:ztf_r_band_exponential_decay_fit}), thus making J0744 a `faint and slow' TDE (section~\ref{sec:discussion}). 

\subsection{\textit{Swift} UVOT}
Follow-up observations with the UVOT instrument on-board \textit{Swift} commenced on MJD$\sim$59164, $\sim$20 days after the eRASS2 detection. A total of 15 observations were then performed over the following 400 day period, using a mix of the $U$, $UVW1$, $UVM2$ and $UVW2$ filters. The level 2 UVOT sky images were again downloaded from the UK Swift Science Data Centre, with aperture photometry being performed on these using the \texttt{uvotsource} task (HEASOFT v6.29, CALDB v20201215). Given the close proximity of the more massive galaxy to the host of J0744, a 3$^{\prime \prime}$ radius source aperture was used, whilst background counts were extracted from a nearby source-free region of radius 15$^{\prime \prime}$. The 3$^{\prime \prime}$ radius was chosen in order to minimise the contamination from the more massive galaxy. Upon visual inspection, it was noticed that a number of image frames were affected by PSF smearing, which was leading to clearly erroneous photometric datapoints; these frames were first identified and discarded before the running of \texttt{uvotsource}. Lastly, an aperture correction was then performed on the coincidence-loss corrected count rates using the \texttt{CURVEOFGROWTH} option within \texttt{uvotsource}, and we also performed the recommended Small Scale Sensitivity check before computing the UVOT photometry \footnote{\url{https://swift.gsfc.nasa.gov/analysis/uvot_digest/sss_check.html}}. In addition to these follow-up observations, the host of J0744 had previously been serendipitously observed by \textit{Swift} UVOT in 2014. We first generated a stack of the three Swift observations performed on 2014-04-29, 2014-11-08 and 2014-11-09, using \texttt{uvotimsum}, and then computed aperture photometry using the same procedure as above (Table~\ref{tab:archival_uvot_photometry}). 
\begin{table}
\centering
\caption{\textit{Swift} UVOT aperture photometry of J0744 obtained from the stacked images of UVOT observations performed in 2014.}
\label{tab:archival_uvot_photometry}
\begin{tabular}{cc}
\hline
Filter & Magnitude \\
\hline
U & 19.66 $\pm$0.06 \\
UVW1 & 21.22 $\pm$0.22 \\
UVM2 & $<22.28$ \\
UVW2 & 22.66 $\pm$0.16 \\
\end{tabular}
\end{table}
\subsection{Photometric lightcurve analysis}
\begin{figure}
    \centering
    \includegraphics[scale=0.75]{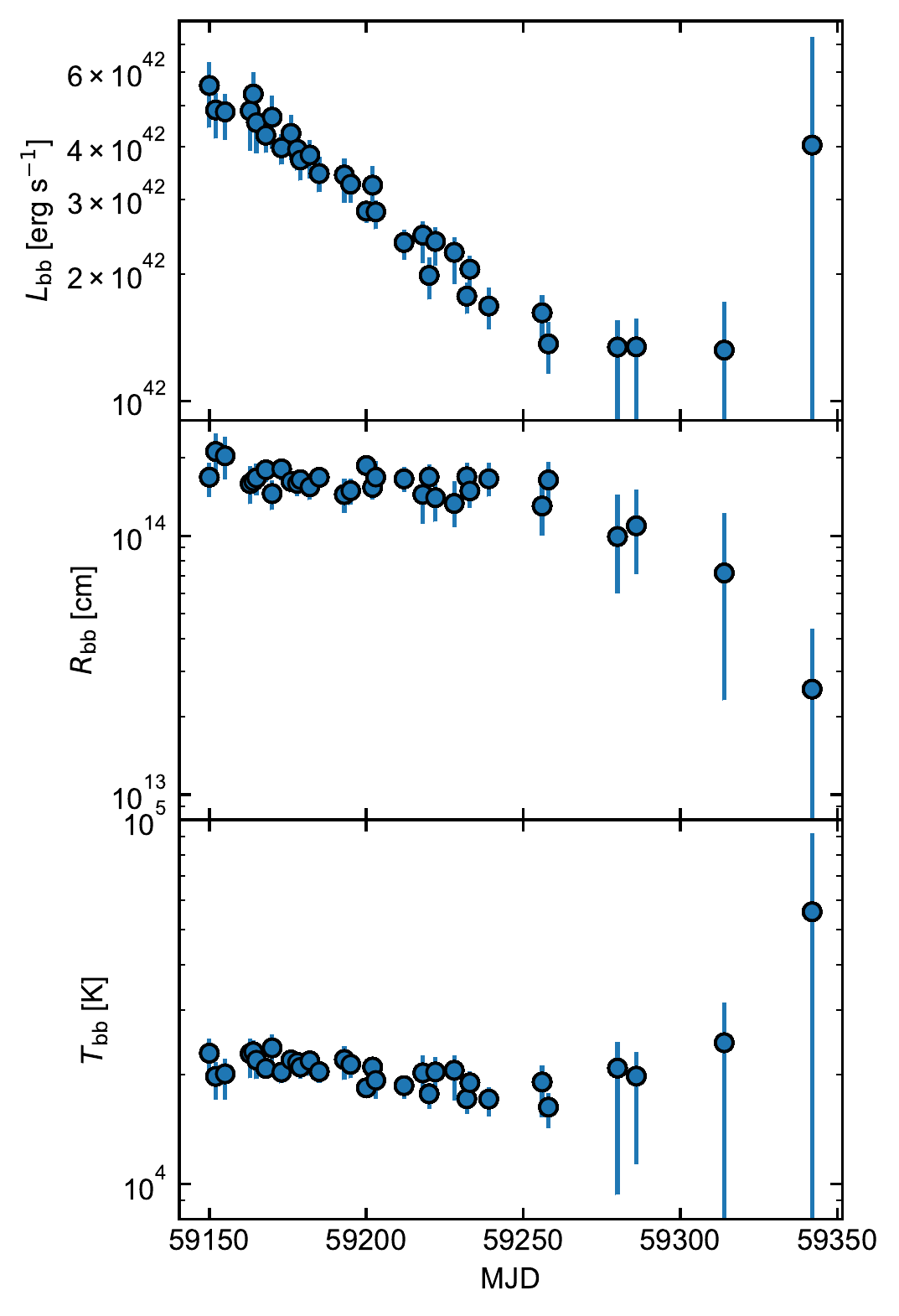}
    \caption{Evolution of blackbody emission inferred from fitting the optical and UV photometry for J0744.}
    \label{fig:optical_bb_evolution}
\end{figure}

For each unique MJD that a \textit{Swift} observation was performed\footnote{This was only done for \textit{Swift} observations obtained between MJD 59150 and 59300, to avoid fitting an SED consisting mainly of extrapolated fluxes.}, we generated synthetic fluxes and errors in the $g$ and $r$ bands using the exponential decay models described in section~\ref{sec:ZTF} (if no ZTF photometry was available for that MJD). Then, the $r$, $g$, $UVW1$, $UVM2$ and $UVW2$ photometry for each MJD was fitted with a blackbody, assuming a log uniform prior on the blackbody temperature, $T_{\mathrm{BB}}$, between $10^{3}$~K and $10^{6}$~K, and on the radius, $R_{\mathrm{BB}}$, between $10^{12}$~cm and $10^{17}$~cm. The evolution of $R_{\mathrm{BB}}$, $T_{\mathrm{BB}}$ and the blackbody luminosity, $L_{\mathrm{BB}}$, inferred via the Stefan-Boltzmann law, is plotted in Fig.~\ref{fig:optical_bb_evolution}. No additional extinction in the host galaxy of J0744 is included here\footnote{From the X-ray spectral fitting (section~\ref{sec:xray_spectral_fitting}), then we do not see evidence for a high neutral hydrogen column density in the host galaxy, with a 2$\sigma$ upper limit on $N_{\mathrm{H}}$ of $6 \times 10^{20}$ cm$^{-2}$ when fitting the eRASS2 spectrum with the \texttt{tbabs*zabspwrlaw} model, equivalent to an $A_{\mathrm{V}}\sim 0.6$ following \citet{predehl_x-raying_1995}. Barring a very high dust-to-gas ratio in the nucleus, we do not consider it likely that there is significant nuclear extinction too.}. As the luminosity decreases over time, the temperature remains approximately constant with $T_{\mathrm{BB}}=(2.1 \pm 0.2)\times 10^4$~K, whilst the radius decreases over time. We also fitted the decay phase of the UV photometry with a power-law of the form $f_{\nu}=[(t-t_0)/\tau]^n$. The inferred $n$ are $-0.9^{+0.2}_{-0.2}$, $-0.83^{+0.17}_{-0.14}$ and $-1.02^{+0.13}_{-0.07}$ for the UVW1, UVM2 and UVW2 bands respectively.
\begin{figure}
    \centering
    \includegraphics[scale=0.75]{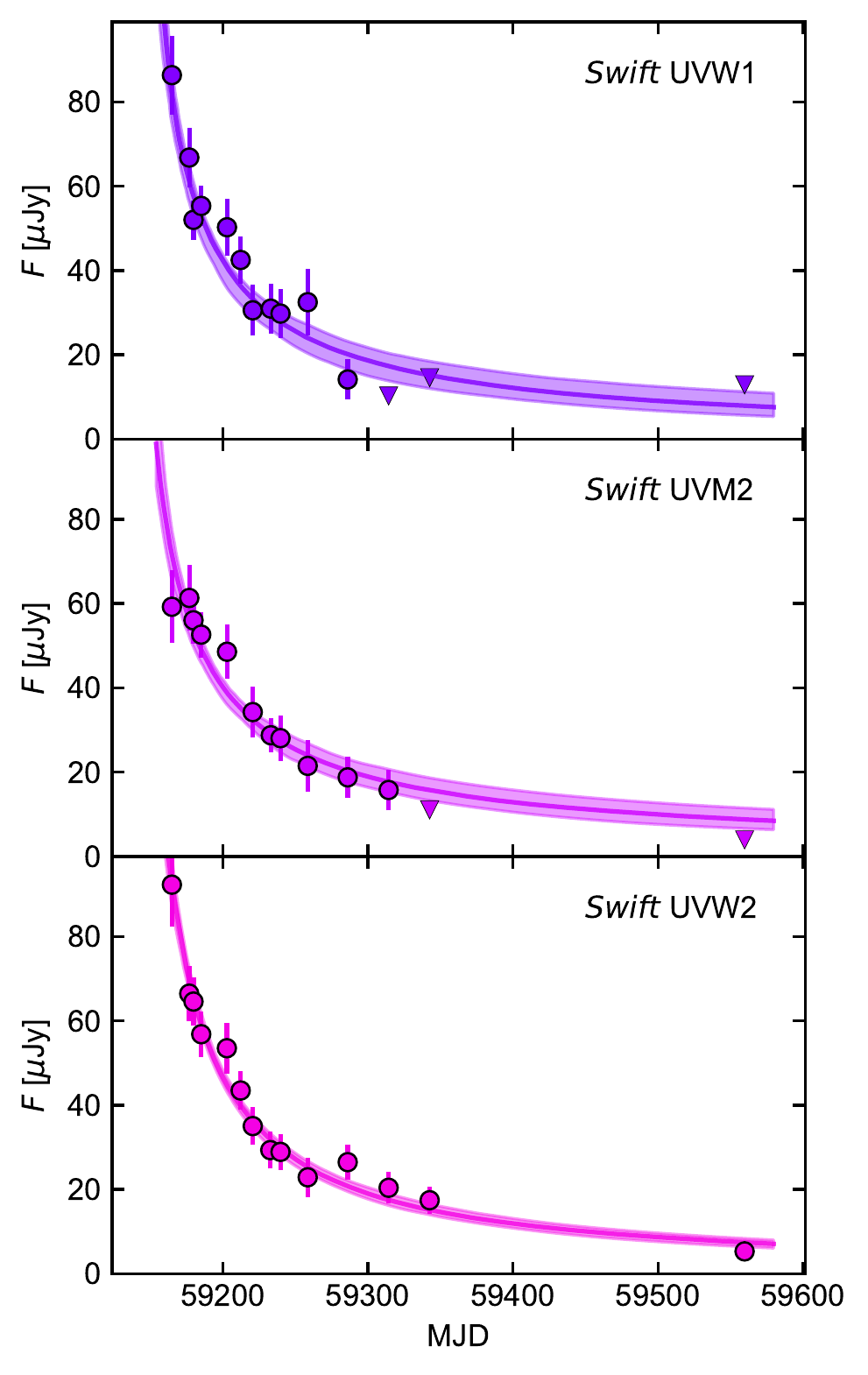}
    \caption{Power-law fits to the \textit{Swift} UV photometry of J0744, with best fitting slopes $-0.9^{+0.2}_{-0.2}$, $-0.83^{+0.17}_{-0.14}$ and $-1.02^{+0.13}_{-0.07}$ for the UVW1, UVM2 and UVW2 bands respectively. Circular and triangle markers represent measured fluxes and 3$\sigma$ ULs respectively, whilst the shaded bands enclose the inner 68\% of the credible region for the fitted model.}
    \label{fig:powerlaw_lightcurve_fits}
\end{figure}

\subsubsection{MOSFiT}
The optical and UV photometry was then fitted using the Modular Open Source Fitter for Transients (MOSFiT; \citealt{guillochon_mosfit_2018}), with the TDE module presented in \citet{mockler_weighing_2019}. Briefly summarising its functionality (see \citealt{mockler_weighing_2019} for a more detailed exposition), this module aims to fit multi-wavelength, multi-epoch photometry with models of the luminosity evolution of TDEs, which have been generated based on simulations \citep{guillochon_hydrodynamical_2013} of the mass fallback rates, $\dot{M}_{\mathrm{fb}}(t)$, of TDEs involving a $1M_{\odot}$ mass star and a black hole with mass, $M_{\mathrm{bh}}=10^6M_{\odot}$, for a range of different impact parameters for the disruption. The $\dot{M}_{\mathrm{fb}}(t)$ are then transformed into viscously-delayed accretion rates, $\dot{M}_{\mathrm{acc}}(t)$, using the viscous timescale for accretion, $T_{\mathrm{viscous}}$, and eq.~7 in \citet{mockler_weighing_2019}. These $\dot{M}_{\mathrm{acc}}(t)$ are then converted into a bolometric luminosity, $L(t)=\epsilon \dot{M}_{\mathrm{acc}}(t)c^2$, where $\epsilon$ is a time-independent accretion efficiency. This is assumed to be reprocessed by material within the vicinity of the black hole, and emitted from a quasi-spherical photosphere with effective temperature:
\begin{equation}
T_{\mathrm{eff}}=\left( \frac{L(t)}{4 \pi \sigma _{\mathrm{SB}} R^2_{\mathrm{phot}}} \right) ^{1/4},
\end{equation}
with $R_{\mathrm{phot}}$ being the radius of the photosphere, defined as:
\begin{equation}
R_{\mathrm{phot}}(t)=R_{\mathrm{ph, 0}}a_{\mathrm{p}}\left( \frac{L(t)}{L_{\mathrm{Edd}}}\right)^{l_{\mathrm{ph}}}
\end{equation}
where $R_{\mathrm{ph, 0}}$ is a unitless normalising factor, $a_{\mathrm{p}}$ is the semi-major axis of the bound stellar debris at peak $\dot{M}_{\mathrm{fb}}(t)$, $L_{\mathrm{Edd}}$ is the Eddington luminosity of the black hole, and $l_{\mathrm{ph}}$ is the photosphere exponent linking $R_{\mathrm{phot}}(t)$ and $L(t)$.

\begin{figure}
    \centering
    \includegraphics[scale=0.7]{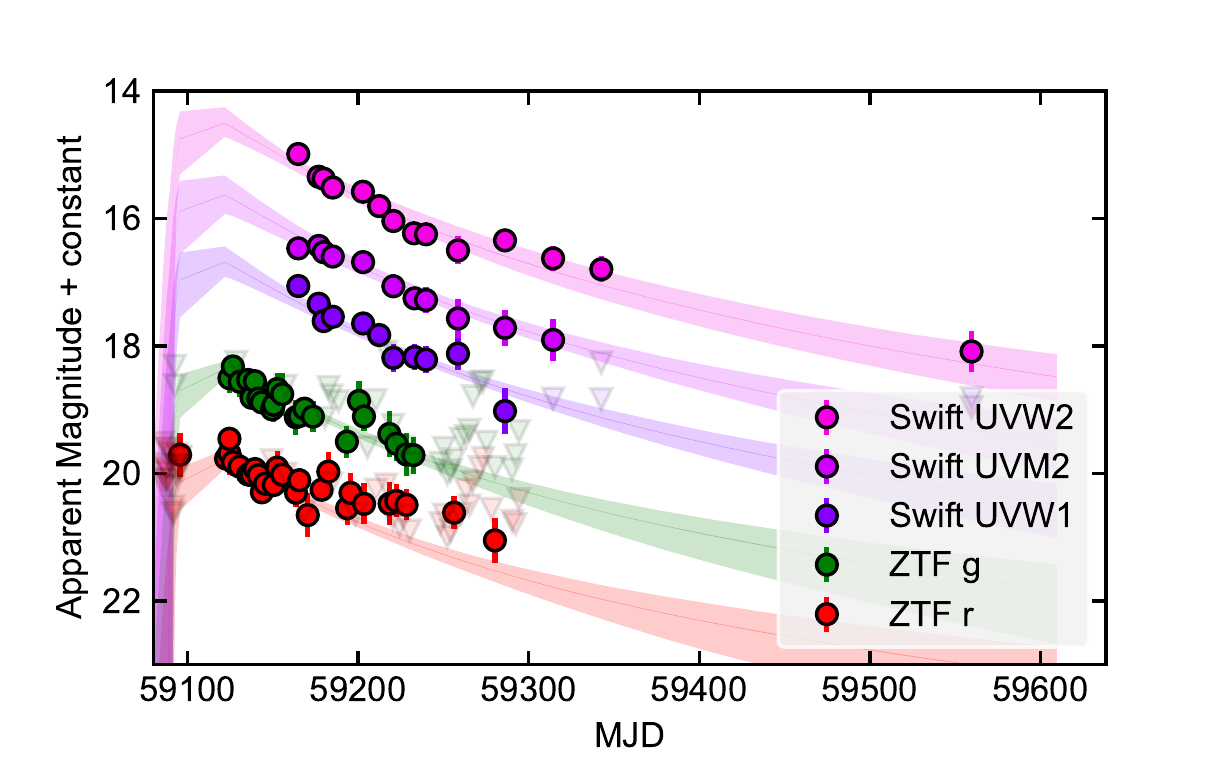}
    \caption{MOSFiT TDE model fits to the ZTF and \textit{Swift} UVOT photometry of J0744. The markers and filled in regions are the same as for Fig.~\ref{fig:powerlaw_lightcurve_fits}. }
    \label{fig:mosfits}
\end{figure}

The free parameters of this model are $M_{\mathrm{bh}}$, $M_{\star}$, $b$ (the scaled impact parameter for the TDE), $\epsilon$, $T_{\mathrm{viscous}}$, $t_{0}$ (the time of first debris fallback measured relative to peak luminosity), $R_{\mathrm{ph, 0}}$, $l_{\mathrm{ph}}$, and $n_{\mathrm{H, host}}$ (the neutral hydrogen column density in the TDEs host galaxy). We adopt the same priors on each parameter as those reported in \citet{mockler_weighing_2019}, with the exception of the prior on the $t_{0}$, which we set to a uniform prior within $\pm50$~days of peak optical luminosity, and $M_{\star}$, for which we use a log-uniform prior between 0.01~$M_{\odot}$ and 100~$M_{\odot}$. Fits were run using the nested sampler \texttt{dynesty} \citep{speagle_dynesty_2020} until convergence, and we performed a simple check of our parameter estimates through ensuring that two successive, differently-seeded MOSFiT runs returned consistent values. The fitted MOSFiT TDE model (Fig.~\ref{fig:mosfits}) reproduces both the early and late-time observed fluxes in the UVW1, UVM2 and UVW2 filters, but the fit to the ZTF $r$-band data underestimates the flux from $\sim 80$ days after the optical peak. Similar excesses between model and data have been seen in the MOSFiT modelling of other TDEs in the literature \citep{mockler_weighing_2019,nicholl_systematic_2022}, suggesting that single-component models for the emission are likely not capable of capturing the full optical-UV evolution of TDEs (potentially due to an additional contribution from the disc emission). 

A full list of parameter estimates generated from the posteriors is presented in Table~\ref{tab:mosfit_stats}.  The inferred black hole mass is $\log (M_{\mathrm{bh}}/ M_{\odot})=6.01 ^{+0.29}_{-0.35}$, whilst the stellar mass is $M_{\star}/ M_{\odot}=(0.09 ^{+0.03}_{-0.02}$)$\pm$0.66 (i.e. unconstrained and dominated by systematic error here). The scaled impact parameter is constrained to be $0.50 ^{+0.25}_{-0.16}$, suggesting that the TDE was produced by a partial, instead of full, disruption. The peak bolometric luminosity, produced by the reprocessing component, is $\sim 2 \times 10^{43}$ erg~s$^{-1}$, and the integrated emitted energy from this is $\sim 2 \times 10^{50}$~erg, corresponding to a total accreted mass of $M_{\mathrm{acc}}\sim 0.002 \, M_{\odot} \left( \epsilon / 0.05 \right)$ (Table~\ref{tab:mosfit_stats}). It is important to consider that these estimates were inferred from MOSFiT fitting a single temperature blackbody model to the UV photometry, and does not model the contribution from the accretion disc in the far-UV to soft X-rays. Whilst such an approach may work for some optically-selected TDEs that have their early time optical emission dominated by stream-stream collisions (e.g.~\citealt{piran_disk_2015}), it is not currently clear how MOSFiT's parameter estimates are systematically biased through not directly modelling the contribution from the disc to the SED. 
\begin{table}
\centering
\caption{Posterior medians and 1$\sigma$ credible regions inferred from the MOSFiT TDE lightcurve fitting. The estimated systematic errors on each estimate are taken from \citet{mockler_weighing_2019}.}
\label{tab:mosfit_stats}
\begin{tabular}{ccc}
\hline
Parameter & Value & Systematic Error\\
\hline
$\log (M_{\mathrm{bh}}/ M_{\odot})$ & $6.01 ^{+0.29}_{-0.35}$ & $\pm$0.2 \\
$ M_{\star}/ M_{\odot}$ & $0.09 ^{+0.03}_{-0.02}$ & $\pm$0.66 \\
$b$ & $0.50 ^{+0.25}_{-0.16}$ & $\pm$0.35 \\
$\log ( \epsilon )$ & $-1.30 ^{+0.28}_{-0.37}$ & $\pm$0.68 \\
$\log (R_{\mathrm{ph, 0}})$ & $-0.14 ^{+0.22}_{-0.18}$ & $\pm$0.4 \\
$l_{\mathrm{ph}}$ & $0.32 ^{+0.28}_{-0.09}$ & $\pm$0.2 \\
$\log (T_{\mathrm{viscous}} / \mathrm{days})$ & $-1.61 ^{+0.94}_{-0.84}$ & $\pm$0.1 \\
$t_{\mathrm{0}}$ (days) & $-7.70 ^{+1.41}_{-2.20}$ & $\pm$15
\end{tabular}
\end{table}

\section{Host galaxy properties}\label{sec:host_properties}
\subsection{Optical spectroscopy}\label{sec:optical_spec}
\subsubsection{Archival}
\begin{figure}
    \centering
    \includegraphics[scale=0.8]{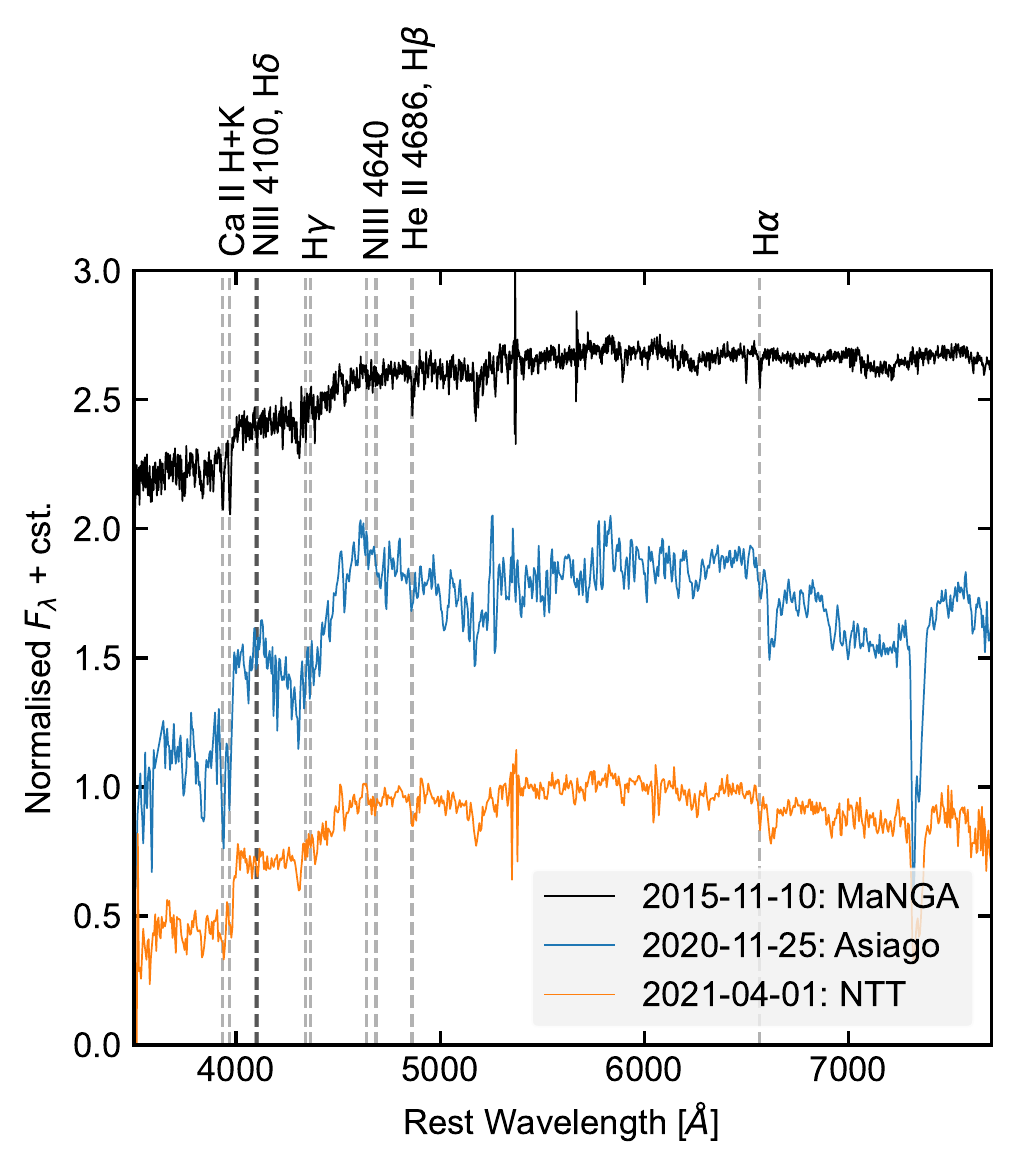}
    \caption{Optical spectra of the host galaxy of J0744, suggesting that the host was quiescent prior to the optical, UV and X-ray outburst observed in late November 2020.}
    \label{fig:optical_spectra}
\end{figure}

The host galaxy of J0744 was observed on 2015-11-10 by the SDSS Mapping Nearby Galaxies at Apache Point Observatory (MaNGA) survey \citep{bundy_overview_2014}, which provides integral field spectroscopy for galaxies with $z < 0.15$. The data cube for this observation (plate-IFU: 8146-3702) was downloaded from the Marvin web application \citep{cherinka_marvin_2019}, and then analysed with the MaNGA Data Reduction Pipeline (DRP; \citealt{westfall_data_2019}), using the latest observation summary file (DRPALL~v3\_1\_1) produced by the DRP. We extracted a spectrum from a 0.5$^{\prime \prime}$x0.5$^{\prime \prime}$ spaxel with spaxel coordinates (32,17), covering the \textit{Gaia} position of J0744, with this showing a quiescent galaxy-like spectrum (Fig.~\ref{fig:optical_spectra} and \ref{fig:ppxf_fit}). Combined with the non-detection in the archival \textit{Chandra} observation (section~\ref{sec:archival_xray}), then it is highly likely that J0744 did not harbour a luminous actively accreting supermassive black hole prior to the 2020 outburst. 

\subsubsection{Follow-up observations}
Several spectroscopic observations of J0744 were performed (Table~\ref{tab:spectroscopic_log}), with the first spectrum being obtained with AFOSC mounted at the 1.82-m telescope (OAPd/INAF) of the Asiago observatory $\sim$52 days after the time of peak observed optical brightness of J0744. It was observed on Nov 25, 2020 and Nov 26, 2020 with VPH7 grism, which covers a range of 3140-7280 \AA\ with a dispersion of about 4.88 \AA/px, and finally on Nov 27, 2020 with VPH6 grism, which convers a range of 4850-9300 \AA\ with a dispersion of about 5.26 \AA/px. The 1.69\arcsec-slit gave a resolution of about 16.5 \AA\ with VPH7 and about 15.5 \AA\ with VPH6. 1$\times$1200 sec exposure was obtained on Nov 25, 2$\times$1200 sec on Nov 26, and 4$\times900$ sec on Nov 27 under a seeing of about 1.5\arcsec, 1.7\arcsec and 1.3\arcsec, respectively.
Spectra were reduced with IRAF following the usual procedure of overscan subtraction, flat-field correction, wavelength calibration by means of mercury-cadmium and neon lamps, flux calibration through the observation of the standard star G191-B2B, and finally night-sky subtraction. The three spectra obtained with grism VPH7 were combined after having overlapped them and checked their agreement. Finally, a total 3500-9300 \AA\ spectrum was obtained by adding the two spectral ranges extracted by choosing an aperture of 4\arcsec.
\begin{table}
	\centering
	\caption{Log of spectroscopic follow-up observations of J0744.}
	\label{tab:spectroscopic_log}
	\begin{tabular}{cccccc} 
		\hline
		MJD & Telescope & Instrument & Exposure [ks] & Airmass \\
		\hline
		59178.138 & Asiago 1.82-m & AFOSC & 1.2 & 1.05 \\ 
		59179.118 & Asiago 1.82-m & AFOSC & 2.4 & 1.05 \\
		59180.107 & Asiago 1.82-m & AFOSC & 3.6 & 1.05 \\
		59305.009 & NTT & EFOSC2 & 2$\times$1.8ks & 1.9 \\ 
		\hline
	\end{tabular}
\end{table}

A second observation was then performed $\sim$4 months later with the ESO Faint Object Spectrograph and Camera (v.2) (EFOSC2; \citealt{buzzoni_eso_1984}) mounted on the ESO New Technology Telescope (NTT) on 2021 March 31 (proposal ID 106.21RU.001, PI: Malyali). We used the grism 13 and the 1.2” slit oriented 90 degrees off the parallactic angle. We obtained two consecutive exposures, 1800 seconds each with the seeing around 1$^{\prime \prime}$ during the observation. The spectrum has the wavelength range of 3685 – 9315 A with a dispersion of 2.77 A/pixel. The data was reduced and calibrated using the \texttt{esoreflex} pipeline (\citealt{freudling_automated_2013}, v2.11.5). The He+Ar arcs were used to obtain the wavelength calibration,  and the standard star LTT3864 was used for flux calibration, which was observed with the same grism and the same slit oriented along the parallactic angle.

\subsubsection{Spectral analysis}
\begin{figure}
    \centering
    \includegraphics[scale=0.8]{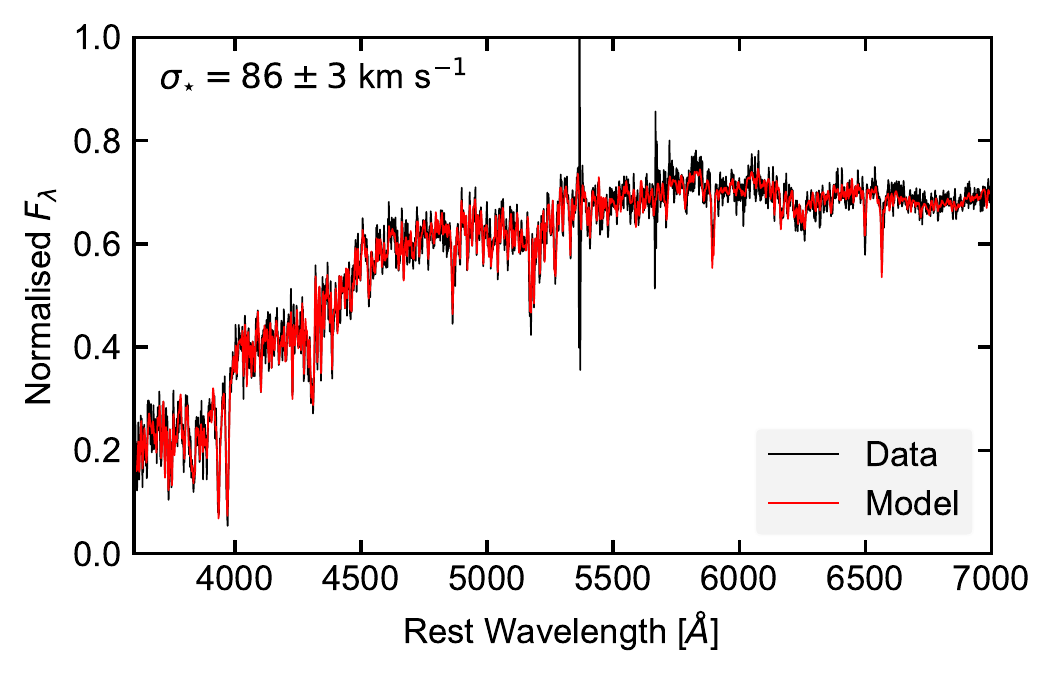}
    \caption{pPXF model fit (red) to the pre-outburst MaNGA spectrum (black) of J0744's host galaxy.}
    \label{fig:ppxf_fit}
\end{figure}

The pre-outburst MaNGA spectrum was first fitted using the penalised pixel fitting routine (pPXF, \citealt{cappellari_parametric_2004,cappellari_improving_2017}), using the stellar template library MILES-HC \citep{westfall_data_2019} generated from the MILES library \citep{falcon-barroso_updated_2011}. The best fitting model is shown in Fig.~\ref{fig:ppxf_fit}, from which we infer a stellar velocity dispersion, $\sigma _{\star}=86 \pm 3$~km~s$^{-1}$ (corrected for instrumental dispersion following the procedure suggested in \citealt{westfall_data_2019}), and $\log [M_{\mathrm{BH}} / M_{\odot}]=6.3 \pm 0.1$ when using the $M_{\mathrm{BH}}-\sigma_{\star}$ relation in \citet{mcconnell_revisiting_2013}.

\begin{figure}
    \centering
    \includegraphics[scale=0.8]{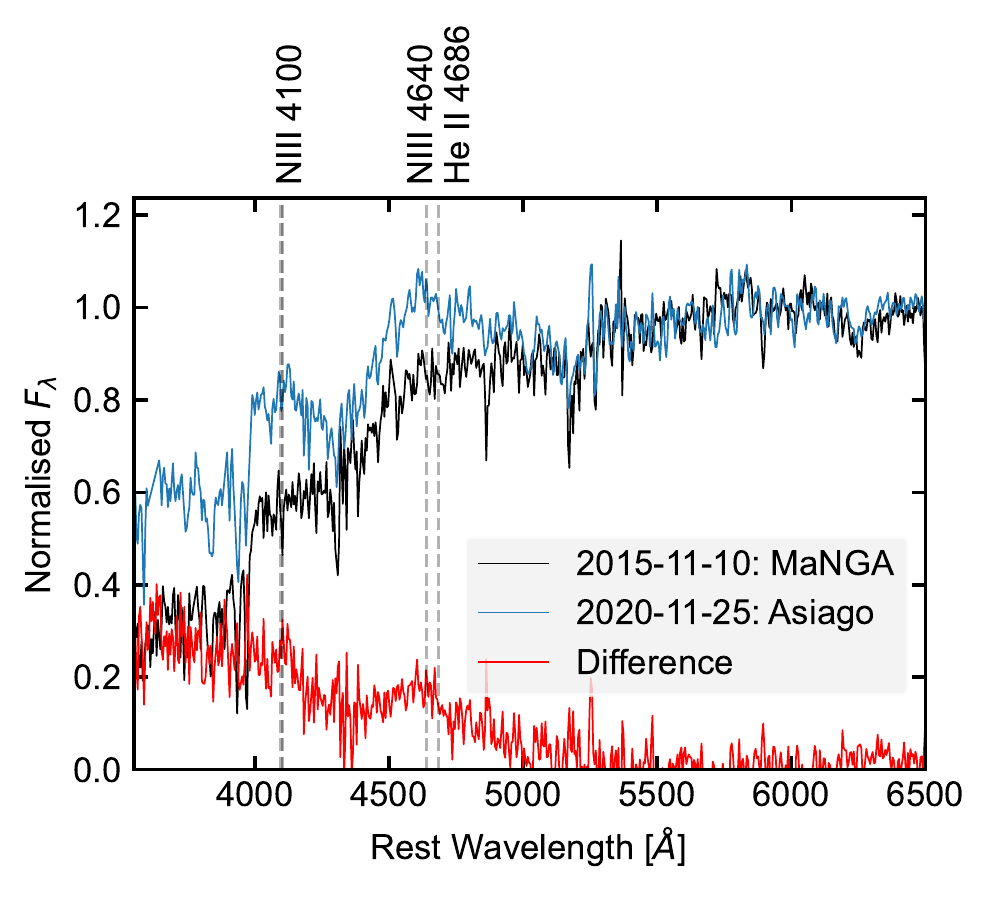}
    \caption{Comparison of the pre-outburst spectrum (MaNGA) with the Asiago spectrum taken $\sim$52~days after peak optical brightness. Both spectra have been normalised by their flux in the 5800-6200\AA~range. The difference between these two spectra is plotted in red, revealing the presence of a transient blue continuum and possible broad emission features around 4100\AA~and 4600\AA~. }
    \label{fig:spectra_difference}
\end{figure}

Each of the follow-up optical spectra of J0744 show a quiescent galaxy-like spectrum, with no `typical' AGN-like emission lines detected above the host emission (Fig.~\ref{fig:optical_spectra}). To search for transient spectral features post-outburst, we downsampled the MaNGA spectrum to the Asiago spectrum, normalised both spectra by the mean of their continuum emission in the 5800-6200\AA~range, and then subtracted the MaNGA from the Asiago spectrum (Fig.~\ref{fig:spectra_difference}). A transient blue continuum is present in the difference spectrum (Fig.~\ref{fig:spectra_difference}), with possible broad emission features being present near 4100\AA~and 4600\AA~. Although the difference spectrum is noisy, we attribute the emission feature near 4600\AA~ to be due to a potential blend of He~II and N~III 4640 \AA~ (highly blueshifted H$\beta$ is disfavoured due to the absence of broad H$\alpha$ emission; Fig.~\ref{fig:optical_spectra}).

\subsection{Morphology}\label{sec:morphology}
One of the most striking aspects of J0744 is its host galaxy (Fig.~\ref{fig:finder_chart}), which is part of a pair of quiescent galaxies at $z=0.0396$. Using the \textit{Gaia} EDR3 \citep{gaia_collaboration_gaia_2016,gaia_collaboration_gaia_2021} positions for each galaxy, the separation between this pair is $\sim 5.4 ^{\prime \prime}$, corresponding to a projected distance of $\sim$4.3~kpc. 
A detailed analysis of their morphologies has previously been performed using SDSS DR7 imaging in \citet{simard_catalog_2011}. In this work, the objects were first deblended using the SExtractor software \citep{bertin_sextractor_1996}, before three different models were fitted to the host: i) a pure S\'ersic decomposition, ii) an exponential disc and de Vaucouleurs bulge (with S\'ersic index fixed to 4), iii) an exponential disc and bulge with S\'ersic index left as a free parameter, with the $F$-statistic then used for selecting the best fitting model amongst these. The inferred morphological parameters for the host of J0744 and its companion are presented in Table~\ref{tab:morphology}, with \citet{simard_catalog_2011} finding that an exponential disc and bulge, with $n_{\mathrm{b}}=3.5\pm0.3$, is the best fitting model for the host of J0744. The inferred half-light radius of the more massive galaxy, $\sim$4~kpc, is comparable to the projected separation the centres of galaxy pair ($\sim$4.3~kpc).

The Galactic extinction corrected SDSS $g$-band model magnitude of the host is $17.89\pm0.01$, translating to an absolute $g$-band magnitude of $-18.39 \pm 0.01$. Using this, along with the half-light radius of J0744's host ($\sim 1.8$~kpc; \citealt{simard_catalog_2011}), and comparing to the population of dwarf galaxies\footnote{The host of J0744 is too extended and too luminous to be a globular cluster.} reported in Fig.~6 of \citet{torrealba_feeble_2016}, one sees that the host of J0744 is at the more luminous, and extended, end of known dwarf galaxies, and has similar luminosity and half-light radius to the Large Magellanic Cloud.

Lastly, we used the relation between $M_{\mathrm{BH}}$ and absolute $r$-band bulge magnitude, $M_{\mathrm{R,\, bulge}}$, presented in equation 5 in \citep{mclure_black_2002}, to infer $\log (M_{\mathrm{BH}} / \mathrm{M}_{\odot})=6.2\pm0.6$ for the disrupting BH in J0744, consistent with $M_{\mathrm{BH}}$ inferred via the $M_{\mathrm{BH}} - \sigma _{\star}$ relation. In doing this, $M_{\mathrm{R,\, bulge}}$ was computed using the SDSS $r$-band model magnitude and the bulge-to-total fraction of 0.51 in \citet{simard_catalog_2011}.

\section{Discussion}\label{sec:discussion}
Due to the combination of the large amplitude, ultra-soft X-ray flare (section~\ref{sec:xray_observations}), the optical variability (section~\ref{sec:photometric_evolution}), and the optical spectrum that shows no strong signs of past AGN activity in the host galaxy (section~\ref{sec:optical_spec}), and their strong similarity to other TDE candidates reported in the literature, then we consider a TDE as the most suitable explanation for the extreme variability seen in this system. Assuming such an interpretation, then the eROSITA detection of ultra-soft X-ray emission within $\sim$20 days of the observed optical peak, which is assumed to be emitted from the innermost regions of the nascent accretion disc, suggests that the disc formed promptly after disruption in this TDE (in contrast to systems that show evidence for delayed disc formation, e.g. \citealt{gezari_x-ray_2017,van_velzen_seventeen_2021,malyali_at_2021,chen_at2019_2022}). 

\subsection{Short timescale, large amplitude X-ray variability in TDEs}\label{sec:xray_variability_in_tdes}
\begin{figure}
    \centering
    \includegraphics[scale=0.8]{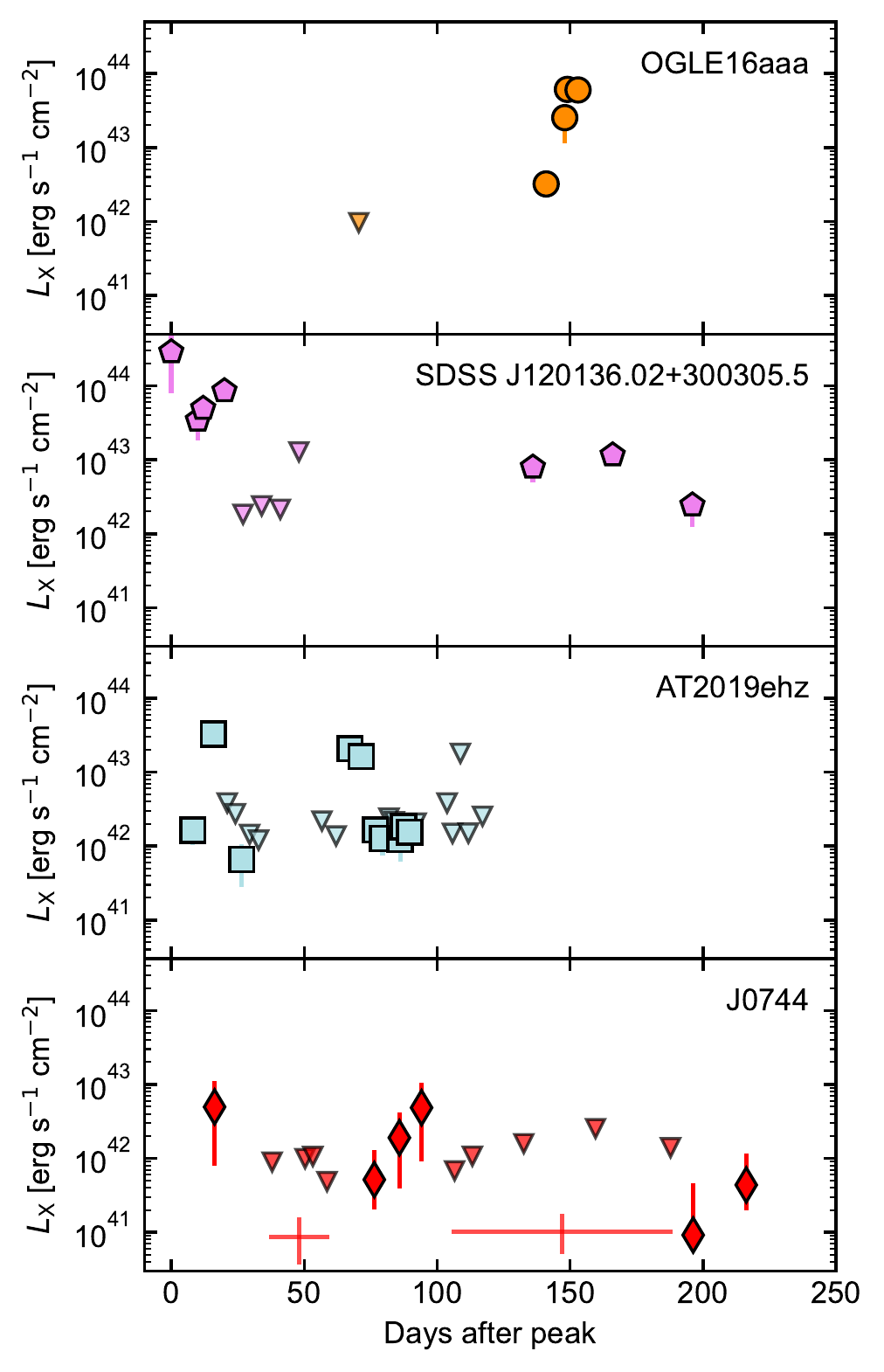}
    \caption{Compilation of soft X-ray lightcurves for TDEs reported in the literature that show large amplitude variability over short timescales, with data taken from \citet{shu_x-ray_2020} for OGLE16aaa, \citet{saxton_tidal_2012} for SDSS J120136/02+300305.5, the \textit{Swift} XRT for AT~2019ehz, with redshift reported in \citet{van_velzen_seventeen_2021}. All luminosities are corrected for Galactic absorption and plotted in the 0.3--2~keV band, except for SDSS J120136.02+300305.5, which is in the 0.2--2~keV band. Triangle markers denote 3$\sigma$ upper limits.}
    \label{fig:tde_xray_lightcurve_comparison}
\end{figure}

The 0.3--2~keV band lightcurve of J0744 shows a fading, and then rebrightening, by a factor $\sim$50 during an 80 day period following the eRASS2 detection (Section~\ref{sec:xray_long_term_variability}, Fig.~\ref{fig:multiwavelength_evolution}). This extreme variability is only seen in the X-rays, with the optical-UV flux continuing its monotonic decline during this period, likely suggesting a different physical origin for these two components. Other similar cases of large amplitude X-ray variability, over short timescales ($\sim$ days), have now also been reported in a number of non-jetted TDEs in the literature (OGLE16aaa, \citealt{kajava_rapid_2020,shu_x-ray_2020}; AT2019ehz, \citealt{van_velzen_seventeen_2021}; SDSS J120136.02+300305.5, \citealt{saxton_tidal_2012}), with their soft-band lightcurves shown together in Fig.~\ref{fig:tde_xray_lightcurve_comparison}. We note that such variability differs from TDEs that show a gradual late time X-ray brightening relative to the optical peak (e.g. increases by a factor of $\sim$10 over a $\sim$200 day period), such as in ASAS-SN 15oi \citep{gezari_x-ray_2017} and AT2019ahz \citep{van_velzen_seventeen_2021,liu_uvoptical_2022,hinkle_discovery_2020,goodwin_at2019azh_2022}. Whilst we do not place any statistical constraints on the prevalence of this short-timescale X-ray variability in TDEs in this work, we highlight here that similar cases of extreme variability may be relatively common in the early stages of X-ray bright TDEs. 

We would disfavour this variability to be due to obscuration of the inner disc by the unbound stellar debris generated in the initial disruption, as this has been predicted to subtend small solid angles measured with respect to the black hole (e.g.~\citealt{krolik_asassn-14li_2016}). For the flaring in OGLE16aaa, an alternate mechanism suggested to explain the variability was that the late-time flare in 0.3--2~keV band was the result of the geometric thinning of a slim disc \citep{wen_continuum-fitting_2020,kajava_rapid_2020}, where, as the disc thinned, then the amount of self-obscuration of its innermost regions would decrease. Again, such an origin would be disfavoured at least for J0744 and AT2019ehz, since these systems are bright at early times, fade, and then rebrighten, which would disagree with the net brightening predicted under the thinning disc scenario. 

Other possible mechanisms for driving such extreme variability fall into three main categories, consisting of the TDE occuring in a supermassive black hole binary, inhomogeneous accretion flows, and the reprocessing of the X-ray emission from the disc. The former has previously been suggested in the literature to explain the lightcurves of OGLE16aaa \citep{shu_x-ray_2020} and SDSS~J120136.02+300305 \citep{liu_milliparsec_2014}, where the authors suggest that the presence of the secondary black hole in the binary interrupts the accretion flow onto the primary, causing large amplitude dips in the soft X-ray flux. 

Alternatively, these lightcurves may be a by-product of the extreme nature of the accretion flows formed in TDEs, with the non-steady-state `discs' fluctuating in effective temperature and observed 0.3--2~keV flux. If this is the case, then a timing analysis of the X-ray lightcurves could be used as a signature to distinguish TDEs from non-TDE induced accretion flares in future studies. One caveat to such an origin, as also noted in \citet{van_velzen_seventeen_2021}, is that for both J0744 and AT~2019ehz, the 0.3--2~keV flux fades, and then rebrightens back to a flux consistent with the previous peak observed flux. It is not immediately clear why such a rebrightening behaviour is observed, if the X-ray variability is produced by random fluctuations in the disc temperature (e.g.~\citealt{mummery_high-energy_2022}). 

One potential avenue to explain this latter point would be through assuming that the disc's luminosity initially remains approximately constant over time, such as in an Eddington-limited plateau phase, but the amount of reprocessing of the disc's X-ray emission along the observer's line of sight varies over time. The exact nature of such a reprocessor is currently unclear, but could potentially be a disc wind (e.g.~\citealt{metzger_bright_2016,dai_unified_2018}), outflows launched from stream-stream collisions \citep{lu_formation_2020}, or absorption by a gaseous envelope surrounding the TDE \citep{loeb_optical_1997}. If the envelope provided only a partial covering of the black hole (e.g. was clumpy), then the rebrightening episodes would thus coincide with the observer momentarily seeing the innermost regions of the accretion disc through the gaps in this envelope.  

\subsection{`Faint and slow' TDEs}\label{sec:faint_and_slow}
A number of optically-selected TDEs in the literature have now been dubbed ‘faint and fast’ TDEs. The prototype of this class, iPTF-16fnl \citep{blagorodnova_iptf16fnl_2017}, was initially identified as an outlier relative to the observed optically-bright TDE population, with respect to both its peak optical luminosity ($(1\pm0.15) \times 10^{43}$~erg~s$^{-1}$, around an order of magnitude fainter than other TDEs), and short rise and decay timescales in its lightcurve (exponential decay timescale $\sim$15~days). Additional members of this class were later suggested through observations of AT~2019qiz \citep{nicholl_outflow_2020} and AT2018ahl/ATLAS18mlw \citep{hinkle_scat_2022}. This set of `faint and fast’ TDEs, considered within the context of the broader population of optically-selected TDEs, has motivated the suggestion that TDEs with lower peak optical luminosities will show faster rise and decay timescales in their lightcurves (e.g. \citealt{blagorodnova_iptf16fnl_2017,hinkle_examining_2020}). It is currently unclear why these `faint and fast' events seem to evolve differently to other known TDEs, although it is important to note that these host lower mass SMBHs (around $10^{6}M_{\odot}$), and the mass fallback timescale scales as $t_{\mathrm{fb}} \propto M_{\mathrm{BH}}^{1/2}$. It has also been suggested that these may be due partial TDEs, which may produce mass fallback rates that decline steeper than the `canonical' $t^{-5/3}$ rate \citep{guillochon_hydrodynamical_2013,coughlin_partial_2019,ryu_tidal_2020-3}.

Similarly to the `faint and fast’ TDEs, J0744 shows i) a short rise timescale in its optical lightcurve, ii) a faint peak optical luminosity, and iii) is also hosted by a low mass SMBH. However, its optical lightcurve decays over much longer timescales (section~\ref{sec:photometric_evolution}), and with a peak observed 0.3--2~keV luminosity of $5\times10^{43}$~erg~s$^{-1}$, J0744 is also $\sim2\times 10^{4}$ and $\sim 10^{3}$ brighter than iPTF-16fnl ($\sim 2 \times 10^{39}$~erg~s$^{-1}$) and AT~2019qiz ($\sim 5 \times 10^{40}$~erg~s$^{-1}$)), respectively. The X-ray spectra of J0744 are also ultra-soft (section~\ref{sec:xray_spectral_fitting}), contrasting with the hard spectrum of AT~2019qiz  ($\Gamma=1.1^{+0.6}_{-0.4}$; \citealt{nicholl_outflow_2020}) of uncertain origin, since it is clearly distinct from the X-ray spectra of other thermal TDEs. J0744 thus shows direct evidence for prompt disc formation only $\sim$20 days after optical peak (again assuming that the ultra-soft X-ray emission originates from the newly formed disc), whereas AT~2019qiz and iPTF-16fnl only show indirect evidence via the transient Bowen emission lines in their optical spectra; the presence of these lines has previously been attributed to the signature of obscured accretion \citep{leloudas_spectral_2019} when observed in the absence of FUV/ X-rays from the accretion disc\footnote{This stems from the Bowen fluorescence mechanism being driven by He~II Lyman $\alpha$ photons, with such emission lines being produced by photons with energy above the ionisation potential of He~II (>54.4~eV).}. We note that we classify the \textit{optical} lightcurve of J0744 as showing ‘faint and slow’ behaviour. Given the uncertainty associated with the bolometric correction here, then it is not clear that this optical faintness necessarily extends to the bolometric luminosity. For instance, \citet{mummery_unified_2021} find that disc-dominated TDEs radiate only a fraction ($\sim$1\% ) of the disc’s luminosity in the optical/ near-UV and X-ray bands, with the majority instead being released in the FUV (see also \citealt{zanazzi_eccentric_2020}). This may be important to consider in any future works modelling the multi-wavelength evolution of J0744-like events.

Following \citet{metzger_bright_2016}, then the ‘faint and slow’ nature of J0744 may potentially be caused by photon trapping in an ionised outflow. If the optical depth to electron scattering is sufficiently high, then the photon diffusion timescale may be less than the expansion timescale for the wind, leading to photons being ‘trapped’ in such an outflow. Photons then advect outwards up to the trapping radius, defined as the radius where the diffusion timescale equals the expansion timescale. During the advection phase, then the temperature decreases adiabatically. For systems with $M_{\mathrm{BH}}\lesssim 10^{7}M_{\odot}$, then adiabatic losses may also cause a suppression of the peak optical luminosity, and also a slowed optical lightcurve decay at early times \citep{metzger_bright_2016}. If X-ray selected TDEs are systematically observed with viewing angles similar to J0744, where trapping effects are important for evolution of the optical luminosity, then X-ray selected/ bright TDEs may show slower decay timescales on average than populations of X-ray dim TDEs. 

Alternatively, given that the most prominent difference between J0744 and other `faint and fast' TDEs is the detection of luminous X-ray emission in J0744, then the differences in the optical lightcurve evolution may be linked to the circularisation efficiency in each TDE. Given the prompt disc formation in J0744, then it is likely that there was efficient circularisation of the stellar debris, potentially akin to the runaway circularisation seen in recent numerical simulations \citep{steinberg_origins_2022}. The optical rise of J0744 may be powered by the circularisation process, whilst the optical decay may be dominated by accretion, either directly from the disc or reprocessed emission (as discussed above). In these systems, the ultra-soft X-ray emission from the newly formed disc may still be affected by time-variable reprocessing (e.g. neutral absorption, or trapping in ionised outflows), hence may show a highly non-monotonic decline (section~\ref{sec:xray_variability_in_tdes}). These systems may also still appear X-ray dim in the first months to years after disruption depending on the observer’s viewing angle \citep{dai_unified_2018}. 

The corollary is that there would be slowed disc formation and inefficient circularisation in `faint and fast' TDEs, where the initial debris may instead settle into an elliptical disc during its infancy. The optical transient emission for these events would then be powered predominantly by outflows launched during the circularisation process \citep{nicholl_outflow_2020}, instead of accretion. The circularisation efficiency of the debris may increase over time (e.g. \citealt{steinberg_origins_2022}), and a fraction of the ‘faint and fast’ TDEs may also show a delayed brightening in the soft X-rays as a result of the slowed accretion in these systems. Although again sensitive to reprocessing and disc physics (e.g. the disc's SED and Comptonisation;~\citealt{mummery_hard_2021}), the late-time X-ray evolution of these systems may appear similar to the decades-long TDE candidate 3XMM J150052.0+015452 (J1500) reported in \citet{lin_likely_2017}. As there was not deep, high-cadence optical photometry monitoring J1500 in the years before its X-ray flaring, then an optical transient associated with that event may easily have been missed. A caveat of this scenario is explaining how the Bowen fluorescence lines are produced in `faint and fast' TDEs if the circularisation is inefficient (although beyond the scope of this work, we note that this may be potentially alleviated by X-rays produced by compressional shocks near pericentre e.g.~\citealt{zanazzi_eccentric_2020,steinberg_origins_2022}). 

\section{Summary}\label{sec:summary}
We have reported on a set of multi-wavelength observations of the TDE candidate eRASSt~J074426.3+291606, located in the nucleus of a quiescent galaxy at $z=0.0396$. The main observed features are as follows:
\begin{enumerate}
    \item J0744 was detected by eROSITA in eRASS2, where it had brightened in the 0.3--2~keV band by a factor $\gtrsim$160 relative to an archival 3$\sigma$ upper limit inferred from \textit{Chandra} observations. The peak observed 0.3--2~keV luminosity is $\sim 5\times10^{43}$~erg~s$^{-1}$, the X-ray spectrum in eRASS2 is ultra-soft ($\Gamma \sim 4$), and remains soft throughout the $\sim$400 day monitoring campaign.
    \item eROSITA, \textit{NICER} XTI and \textit{Swift} XRT observations reveal a net decline in the 0.3--2~keV X-ray flux over a $\sim$400 day monitoring campaign by at least a factor of 10.
    \item The 0.3--2~keV band flux fades, and rebrightens, by a factor $\sim$50 over an 80 day period, deviating significantly from a smooth $t^{-5/3}$ decay traditionally expected for TDEs. We considered a set of large amplitude X-ray variability in TDEs, and suggest that such variability may not be uncommon in the early stages of X-ray bright TDE lightcurves.  This non-monotonic declining behaviour will be important to consider in future work computing the X-ray luminosity functions of TDEs (e.g. when inferring the peak X-ray luminosity from an infrequently sampled X-ray lightcurve).
    \item J0744 is accompanied by an optical-UV flare, which localises this transient to the nucleus of the quiescent galaxy SDSS~J074426.12+291607.4. The detection of ultra-soft X-rays only 20 days after peak optical brightness suggests the prompt formation of an accretion disc in this TDE. Although this optical flare was detected by ZTF, J0744 was missed out from the recent sample of ZTF-selected TDEs \citep{van_velzen_seventeen_2021,hammerstein_final_2022}.
    \item The peak optical luminosity of J0744 is extremely faint (peak absolute $g$-band magnitude $\sim -16.8$ mag). Whilst the optical $g$-band emission makes J0744 the faintest TDE observed to date, it is not `faint and fast' like other optically-selected TDEs of similar peak luminosity; instead, it is `faint and slow' (section~\ref{sec:faint_and_slow}). 
\end{enumerate}

Higher redshift variants of this system, or for systems where the host galaxy is even fainter relative to the more massive nearby galaxy than in J0744, may appear as off-nuclear TDEs in future surveys. To date, J0744 is the second strong, off-nuclear TDE candidate reported in the literature (3XMM~J215022.4-055108; \citealt{lin_luminous_2018}), with this system also being X-ray bright. X-ray and UV surveys may be particularly well suited to identifying IMBHs in galaxies, benefiting from the lower amount of contamination from other transient classes (such as supernovae), and due to the transient UV emission being relatively long lived and easier to detect over the quiescent host than at optical wavelengths. Similar events to J0744 will likely be detected by future planned missions, such as with the \textit{Einstein Probe} (\textit{EP}; \citealt{yuan_einstein_2018}), which will also provide high-cadence monitoring capabilities for events over the 0.3--10~keV range. In the UV, then TDEs similar to J0744 may be exceptionally well studied with the \textit{Ultraviolet Transient Astronomy Satellite} (\textit{ULTRASAT}; \citealt{sagiv_science_2014}), scheduled for launch in 2025, and later also with the \textit{Ultraviolet Explorer} (\textit{UVEX}; \citealt{kulkarni_science_2021}). These future planned observatories may therefore prove an invaluable tool to uncover TDEs occuring in BHs with masses $\lesssim 10^{6}M_{\odot}$, and for probing the black hole occupation fraction at the low mass end.

\section*{Acknowledgements}
AM is grateful to both the \textit{Swift} and \textit{NICER} teams for approving the many ToO requests, and the \textit{Swift} team for their assistance with reducing the UVOT data, in particular to Peter Brown. AM would also like to thank Matt Nicholl for assistance with using MOSFiT. AM acknowledges support by DLR under the grant 50 QR 2110 (XMM\_NuTra, PI: Z. Liu). We would like to thank the referee for a constructive report that improved the quality of the paper.

This work is based on data from eROSITA, the soft X-ray instrument aboard SRG, a joint Russian-German science mission supported by the Russian Space Agency (Roskosmos), in the interests of the Russian Academy of Sciences represented by its Space Research Institute (IKI), and the Deutsches Zentrum für Luft- und Raumfahrt (DLR). The SRG spacecraft was built by Lavochkin Association (NPOL) and its subcontractors, and is operated by NPOL with support from the Max Planck Institute for Extraterrestrial Physics (MPE).

The development and construction of the eROSITA X-ray instrument was led by MPE, with contributions from the Dr. Karl Remeis Observatory Bamberg \& ECAP (FAU Erlangen-Nuernberg), the University of Hamburg Observatory, the Leibniz Institute for Astrophysics Potsdam (AIP), and the Institute for Astronomy and Astrophysics of the University of Tübingen, with the support of DLR and the Max Planck Society. The Argelander Institute for Astronomy of the University of Bonn and the Ludwig Maximilians Universität Munich also participated in the science preparation for eROSITA.

The eROSITA data shown here were processed using the eSASS software system developed by the German eROSITA consortium.

This work made use of data supplied by the UK Swift Science Data Centre at the University of Leicester.

The ZTF forced-photometry service was funded under the Heising-Simons Foundation grant \#12540303 (PI: Graham).

DH acknowledges support from DLR grant FKZ 50 OR
2003. MK is supported by DFG grant KR 3338/4-1.

\section*{Data Availability}
The eRASS1-4 data taken within the German half of the eROSITA sky is currently planned to be made public by Q2 2024. The Swift data is available to download through the UK Swift Data Science website\footnote{\url{https://www.swift.ac.uk/archive/index.php}}, whilst the NICER data is accessible through NASA’s HEASARC interface\footnote{\url{https://heasarc.gsfc.nasa.gov/docs/nicer/nicer_archive.html}}. Publicly available ZTF data can be accessed through the ZTF forced photometry service\footnote{\url{https://irsa.ipac.caltech.edu/Missions/ztf.html}}. Follow-up optical spectra will likely remain private at least until the release of the forthcoming eROSITA-selected TDE population paper, but could be made available upon reasonable request.



\bibliographystyle{mnras}
\bibliography{J0744} 

\begin{thebibliography}{}
\makeatletter
\relax
\def\mn@urlcharsother{\let\do\@makeother \do\$\do\&\do\#\do\^\do\_\do\%\do\~}
\def\mn@doi{\begingroup\mn@urlcharsother \@ifnextchar [ {\mn@doi@}
  {\mn@doi@[]}}
\def\mn@doi@[#1]#2{\def\@tempa{#1}\ifx\@tempa\@empty \href
  {http://dx.doi.org/#2} {doi:#2}\else \href {http://dx.doi.org/#2} {#1}\fi
  \endgroup}
\def\mn@eprint#1#2{\mn@eprint@#1:#2::\@nil}
\def\mn@eprint@arXiv#1{\href {http://arxiv.org/abs/#1} {{\tt arXiv:#1}}}
\def\mn@eprint@dblp#1{\href {http://dblp.uni-trier.de/rec/bibtex/#1.xml}
  {dblp:#1}}
\def\mn@eprint@#1:#2:#3:#4\@nil{\def\@tempa {#1}\def\@tempb {#2}\def\@tempc
  {#3}\ifx \@tempc \@empty \let \@tempc \@tempb \let \@tempb \@tempa \fi \ifx
  \@tempb \@empty \def\@tempb {arXiv}\fi \@ifundefined
  {mn@eprint@\@tempb}{\@tempb:\@tempc}{\expandafter \expandafter \csname
  mn@eprint@\@tempb\endcsname \expandafter{\@tempc}}}

\bibitem[\protect\citeauthoryear{Arcavi et~al.,}{Arcavi
  et~al.}{2014}]{arcavi_continuum_2014}
Arcavi I.,  et~al., 2014, \mn@doi [The Astrophysical Journal]
  {10.1088/0004-637X/793/1/38}, 793, 38

\bibitem[\protect\citeauthoryear{Arnaud}{Arnaud}{1996}]{arnaud_xspec_1996}
Arnaud K.~A.,  1996, in Jacoby G.~H.,  Barnes J.,  eds,  Astronomical {Society}
  of the {Pacific} {Conference} {Series} Vol. 101, Astronomical {Data}
  {Analysis} {Software} and {Systems} {V}. p.~17

\bibitem[\protect\citeauthoryear{Bade, Komossa  \& Dahlem}{Bade
  et~al.}{1996}]{bade_detection_1996}
Bade N.,  Komossa S.,   Dahlem M.,  1996, åp, 309, L35

\bibitem[\protect\citeauthoryear{Bellm et~al.,}{Bellm
  et~al.}{2019}]{bellm_zwicky_2019}
Bellm E.~C.,  et~al., 2019, \mn@doi [Publications of the Astronomical Society
  of the Pacific] {10.1088/1538-3873/aaecbe}, 131, 018002

\bibitem[\protect\citeauthoryear{Bertin \& Arnouts}{Bertin \&
  Arnouts}{1996}]{bertin_sextractor_1996}
Bertin E.,  Arnouts S.,  1996, \mn@doi [Astronomy and Astrophysics Supplement
  Series] {10.1051/aas:1996164}, 117, 393

\bibitem[\protect\citeauthoryear{Blagorodnova et~al.,}{Blagorodnova
  et~al.}{2017}]{blagorodnova_iptf16fnl_2017}
Blagorodnova N.,  et~al., 2017, \mn@doi [The Astrophysical Journal]
  {10.3847/1538-4357/aa7579}, 844, 46

\bibitem[\protect\citeauthoryear{Bonnerot \& Stone}{Bonnerot \&
  Stone}{2020}]{bonnerot_formation_2020}
Bonnerot C.,  Stone N.,  2020, arXiv:2008.11731 [astro-ph]

\bibitem[\protect\citeauthoryear{Brightman et~al.,}{Brightman
  et~al.}{2021}]{brightman_luminous_2021}
Brightman M.,  et~al., 2021, \mn@doi [The Astrophysical Journal]
  {10.3847/1538-4357/abde34}, 909, 102

\bibitem[\protect\citeauthoryear{Brunner et~al.,}{Brunner
  et~al.}{2022}]{brunner_erosita_2022}
Brunner H.,  et~al., 2022, \mn@doi [Astronomy \& Astrophysics]
  {10.1051/0004-6361/202141266}, 661, A1

\bibitem[\protect\citeauthoryear{Buchner}{Buchner}{2021}]{buchner_ultranest_2021}
Buchner J.,  2021, {UltraNest} -- a robust, general purpose {Bayesian}
  inference engine, \url {http://arxiv.org/abs/2101.09604}

\bibitem[\protect\citeauthoryear{Buchner et~al.,}{Buchner
  et~al.}{2014}]{buchner_x-ray_2014}
Buchner J.,  et~al., 2014, \mn@doi [Astronomy \& Astrophysics]
  {10.1051/0004-6361/201322971}, 564, A125

\bibitem[\protect\citeauthoryear{Bundy et~al.,}{Bundy
  et~al.}{2014}]{bundy_overview_2014}
Bundy K.,  et~al., 2014, \mn@doi [The Astrophysical Journal]
  {10.1088/0004-637X/798/1/7}, 798, 7

\bibitem[\protect\citeauthoryear{Burrows et~al.,}{Burrows
  et~al.}{2005}]{burrows_swift_2005}
Burrows D.~N.,  et~al., 2005, \mn@doi [Space Science Reviews]
  {10.1007/s11214-005-5097-2}, 120, 165

\bibitem[\protect\citeauthoryear{Buzzoni et~al.,}{Buzzoni
  et~al.}{1984}]{buzzoni_eso_1984}
Buzzoni B.,  et~al., 1984, The Messenger, 38, 9

\bibitem[\protect\citeauthoryear{Cappellari}{Cappellari}{2017}]{cappellari_improving_2017}
Cappellari M.,  2017, \mn@doi [Monthly Notices of the Royal Astronomical
  Society] {10.1093/mnras/stw3020}, 466, 798

\bibitem[\protect\citeauthoryear{Cappellari \& Emsellem}{Cappellari \&
  Emsellem}{2004}]{cappellari_parametric_2004}
Cappellari M.,  Emsellem E.,  2004, \mn@doi [{\textbackslash}pasp]
  {10.1086/381875}, 116, 138

\bibitem[\protect\citeauthoryear{Cash}{Cash}{1976}]{cash_generation_1976}
Cash W.,  1976, åp, 52, 307

\bibitem[\protect\citeauthoryear{Chen, Dou  \& Shen}{Chen
  et~al.}{2022}]{chen_at2019_2022}
Chen J.-H.,  Dou L.-M.,   Shen R.-F.,  2022, \mn@doi [The Astrophysical
  Journal] {10.3847/1538-4357/ac558d}, 928, 63

\bibitem[\protect\citeauthoryear{Cherinka et~al.,}{Cherinka
  et~al.}{2019}]{cherinka_marvin_2019}
Cherinka B.,  et~al., 2019, \mn@doi [The Astronomical Journal]
  {10.3847/1538-3881/ab2634}, 158, 74

\bibitem[\protect\citeauthoryear{Coughlin \& Nixon}{Coughlin \&
  Nixon}{2019}]{coughlin_partial_2019}
Coughlin E.~R.,  Nixon C.~J.,  2019, \mn@doi [The Astrophysical Journal]
  {10.3847/2041-8213/ab412d}, 883, L17

\bibitem[\protect\citeauthoryear{Dai, McKinney, Roth, Ramirez-Ruiz  \&
  Miller}{Dai et~al.}{2018}]{dai_unified_2018}
Dai L.,  McKinney J.~C.,  Roth N.,  Ramirez-Ruiz E.,   Miller M.~C.,  2018,
  \mn@doi [The Astrophysical Journal] {10.3847/2041-8213/aab429}, 859, L20

\bibitem[\protect\citeauthoryear{Donato et~al.,}{Donato
  et~al.}{2014}]{donato_tidal_2014}
Donato D.,  et~al., 2014, \mn@doi [The Astrophysical Journal]
  {10.1088/0004-637X/781/2/59}, 781, 59

\bibitem[\protect\citeauthoryear{Evans et~al.,}{Evans
  et~al.}{2007}]{evans_online_2007}
Evans P.~A.,  et~al., 2007, \mn@doi [Astronomy \& Astrophysics]
  {10.1051/0004-6361:20077530}, 469, 379

\bibitem[\protect\citeauthoryear{Evans et~al.,}{Evans
  et~al.}{2009}]{evans_methods_2009}
Evans P.~A.,  et~al., 2009, \mn@doi [Monthly Notices of the Royal Astronomical
  Society] {10.1111/j.1365-2966.2009.14913.x}, 397, 1177

\bibitem[\protect\citeauthoryear{Falcón-Barroso, Sánchez-Blázquez, Vazdekis,
  Ricciardelli, Cardiel, Cenarro, Gorgas  \& Peletier}{Falcón-Barroso
  et~al.}{2011}]{falcon-barroso_updated_2011}
Falcón-Barroso J.,  Sánchez-Blázquez P.,  Vazdekis A.,  Ricciardelli E.,
  Cardiel N.,  Cenarro A.~J.,  Gorgas J.,   Peletier R.~F.,  2011, \mn@doi
  [Astronomy \& Astrophysics] {10.1051/0004-6361/201116842}, 532, A95

\bibitem[\protect\citeauthoryear{Freudling, Romaniello, Bramich, Ballester,
  Forchi, García-Dabló, Moehler  \& Neeser}{Freudling
  et~al.}{2013}]{freudling_automated_2013}
Freudling W.,  Romaniello M.,  Bramich D.~M.,  Ballester P.,  Forchi V.,
  García-Dabló C.~E.,  Moehler S.,   Neeser M.~J.,  2013, \mn@doi [Astronomy
  \& Astrophysics] {10.1051/0004-6361/201322494}, 559, A96

\bibitem[\protect\citeauthoryear{{Fruscione} et~al.,}{{Fruscione}
  et~al.}{2006}]{fruscione_ciao_2006}
{Fruscione} A.,  et~al., 2006, in {Silva} D.~R.,  {Doxsey} R.~E.,  eds,
  Society of Photo-Optical Instrumentation Engineers (SPIE) Conference Series
  Vol. 6270, Society of Photo-Optical Instrumentation Engineers (SPIE)
  Conference Series. p. 62701V, \mn@doi{10.1117/12.671760}

\bibitem[\protect\citeauthoryear{Förster et~al.,}{Förster
  et~al.}{2021}]{forster_automatic_2021}
Förster F.,  et~al., 2021, \mn@doi [The Astronomical Journal]
  {10.3847/1538-3881/abe9bc}, 161, 242

\bibitem[\protect\citeauthoryear{{Gaia Collaboration} et~al.,}{{Gaia
  Collaboration} et~al.}{2016}]{gaia_collaboration_gaia_2016}
{Gaia Collaboration} et~al., 2016, \mn@doi [Astronomy \& Astrophysics]
  {10.1051/0004-6361/201629272}, 595, A1

\bibitem[\protect\citeauthoryear{{Gaia Collaboration} et~al.,}{{Gaia
  Collaboration} et~al.}{2021}]{gaia_collaboration_gaia_2021}
{Gaia Collaboration} et~al., 2021, \mn@doi [Astronomy \& Astrophysics]
  {10.1051/0004-6361/202039498}, 649, A6

\bibitem[\protect\citeauthoryear{Gehrels et~al.,}{Gehrels
  et~al.}{2004}]{gehrels_swift_2004}
Gehrels N.,  et~al., 2004, in {AIP} {Conference} {Proceedings}. pp 637--641,
  \mn@doi{10.1063/1.1810924}, \url {http://arxiv.org/abs/astro-ph/0405233}

\bibitem[\protect\citeauthoryear{Gendreau et~al.,}{Gendreau
  et~al.}{2016}]{den_herder_neutron_2016}
Gendreau K.~C.,  et~al., 2016. Edinburgh, United Kingdom, p. 99051H,
  \mn@doi{10.1117/12.2231304}, \url
  {http://proceedings.spiedigitallibrary.org/proceeding.aspx?doi=10.1117/12.2231304}

\bibitem[\protect\citeauthoryear{Gezari et~al.,}{Gezari
  et~al.}{2006}]{gezari_ultraviolet_2006}
Gezari S.,  et~al., 2006, \mn@doi [The Astrophysical Journal] {10.1086/509918},
  653, L25

\bibitem[\protect\citeauthoryear{Gezari et~al.,}{Gezari
  et~al.}{2008}]{gezari_uvoptical_2008}
Gezari S.,  et~al., 2008, \mn@doi [The Astrophysical Journal] {10.1086/529008},
  676, 944

\bibitem[\protect\citeauthoryear{Gezari et~al.,}{Gezari
  et~al.}{2009}]{gezari_luminous_2009}
Gezari S.,  et~al., 2009, \mn@doi [The Astrophysical Journal]
  {10.1088/0004-637X/698/2/1367}, 698, 1367

\bibitem[\protect\citeauthoryear{Gezari, Cenko  \& Arcavi}{Gezari
  et~al.}{2017}]{gezari_x-ray_2017}
Gezari S.,  Cenko S.~B.,   Arcavi I.,  2017, \mn@doi [The Astrophysical
  Journal] {10.3847/2041-8213/aaa0c2}, 851, L47

\bibitem[\protect\citeauthoryear{Goodwin et~al.,}{Goodwin
  et~al.}{2022}]{goodwin_at2019azh_2022}
Goodwin A.~J.,  et~al., 2022, \mn@doi [Monthly Notices of the Royal
  Astronomical Society] {10.1093/mnras/stac333}, 511, 5328

\bibitem[\protect\citeauthoryear{Graham et~al.,}{Graham
  et~al.}{2019}]{graham_zwicky_2019}
Graham M.~J.,  et~al., 2019, \mn@doi [Publications of the Astronomical Society
  of the Pacific] {10.1088/1538-3873/ab006c}, 131, 078001

\bibitem[\protect\citeauthoryear{Greiner, Schwarz, Zharikov  \& Orio}{Greiner
  et~al.}{2000}]{greiner_rx_2000}
Greiner J.,  Schwarz R.,  Zharikov S.,   Orio M.,  2000, åp, 362, L25

\bibitem[\protect\citeauthoryear{Grupe, Thomas  \& Leighly}{Grupe
  et~al.}{1999}]{grupe_rx_1999}
Grupe D.,  Thomas H.~C.,   Leighly K.~M.,  1999, åp, 350, L31

\bibitem[\protect\citeauthoryear{Guillochon \& Ramirez-Ruiz}{Guillochon \&
  Ramirez-Ruiz}{2013}]{guillochon_hydrodynamical_2013}
Guillochon J.,  Ramirez-Ruiz E.,  2013, \mn@doi [The Astrophysical Journal]
  {10.1088/0004-637X/767/1/25}, 767, 25

\bibitem[\protect\citeauthoryear{Guillochon, Nicholl, Villar, Mockler, Narayan,
  Mandel, Berger  \& Williams}{Guillochon
  et~al.}{2018}]{guillochon_mosfit_2018}
Guillochon J.,  Nicholl M.,  Villar V.~A.,  Mockler B.,  Narayan G.,  Mandel
  K.~S.,  Berger E.,   Williams P. K.~G.,  2018, \mn@doi [The Astrophysical
  Journal Supplement Series] {10.3847/1538-4365/aab761}, 236, 6

\bibitem[\protect\citeauthoryear{{HI4PI Collaboration:} et~al.,}{{HI4PI
  Collaboration:} et~al.}{2016}]{hi4pi_collaboration_hi4pi_2016}
{HI4PI Collaboration:} et~al., 2016, \mn@doi [Astronomy \& Astrophysics]
  {10.1051/0004-6361/201629178}, 594, A116

\bibitem[\protect\citeauthoryear{Hammerstein et~al.,}{Hammerstein
  et~al.}{2022}]{hammerstein_final_2022}
Hammerstein E.,  et~al., 2022, \mn@doi [arXiv:2203.01461 [astro-ph]]
  {https://doi.org/10.48550/arXiv.2203.01461}

\bibitem[\protect\citeauthoryear{Hills}{Hills}{1975}]{hills_possible_1975}
Hills J.~G.,  1975, \mn@doi [Nature] {10.1038/254295a0}, 254, 295

\bibitem[\protect\citeauthoryear{Hinkle et~al.,}{Hinkle
  et~al.}{2020a}]{hinkle_discovery_2020}
Hinkle J.~T.,  et~al., 2020a, \mn@doi [Monthly Notices of the Royal
  Astronomical Society] {10.1093/mnras/staa3170}, 500, 1673

\bibitem[\protect\citeauthoryear{Hinkle, Holoien, Shappee, Auchettl, Kochanek,
  Stanek, Payne  \& Thompson}{Hinkle et~al.}{2020b}]{hinkle_examining_2020}
Hinkle J.~T.,  Holoien T. W.-S.,  Shappee B.~J.,  Auchettl K.,  Kochanek C.~S.,
   Stanek K.~Z.,  Payne A.~V.,   Thompson T.~A.,  2020b, \mn@doi [The
  Astrophysical Journal] {10.3847/2041-8213/ab89a2}, 894, L10

\bibitem[\protect\citeauthoryear{Hinkle et~al.,}{Hinkle
  et~al.}{2022}]{hinkle_scat_2022}
Hinkle J.~T.,  et~al., 2022, {SCAT} {Uncovers} {ATLAS}'s {First} {Tidal}
  {Disruption} {Event} {ATLAS18mlw}: {A} {Faint} and {Fast} {TDE} in a
  {Quiescent} {Balmer} {Strong} {Galaxy}, \url
  {http://arxiv.org/abs/2202.05281}

\bibitem[\protect\citeauthoryear{Holoien et~al.,}{Holoien
  et~al.}{2014}]{holoien_asassn-14ae_2014}
Holoien T. W.-S.,  et~al., 2014, \mn@doi [Monthly Notices of the Royal
  Astronomical Society] {10.1093/mnras/stu1922}, 445, 3263

\bibitem[\protect\citeauthoryear{Holoien et~al.,}{Holoien
  et~al.}{2019}]{holoien_ps18kh_2019}
Holoien T. W.-S.,  et~al., 2019, \mn@doi [The Astrophysical Journal]
  {10.3847/1538-4357/ab2ae1}, 880, 120

\bibitem[\protect\citeauthoryear{Jansen et~al.,}{Jansen
  et~al.}{2001}]{jansen_xmm-newton_2001}
Jansen F.,  et~al., 2001, \mn@doi [Astronomy \& Astrophysics]
  {10.1051/0004-6361:20000036}, 365, L1

\bibitem[\protect\citeauthoryear{Kajava, Giustini, Saxton  \& Miniutti}{Kajava
  et~al.}{2020}]{kajava_rapid_2020}
Kajava J. J.~E.,  Giustini M.,  Saxton R.~D.,   Miniutti G.,  2020, \mn@doi
  [Astronomy \& Astrophysics] {10.1051/0004-6361/202038165}, 639, A100

\bibitem[\protect\citeauthoryear{Khabibullin \& Sazonov}{Khabibullin \&
  Sazonov}{2014}]{khabibullin_stellar_2014}
Khabibullin I.,  Sazonov S.,  2014, \mn@doi [Monthly Notices of the Royal
  Astronomical Society] {10.1093/mnras/stu1491}, 444, 1041

\bibitem[\protect\citeauthoryear{Komossa \& Bade}{Komossa \&
  Bade}{1999}]{komossa_giant_1999}
Komossa S.,  Bade N.,  1999, åp, 343, 775

\bibitem[\protect\citeauthoryear{Komossa \& Greiner}{Komossa \&
  Greiner}{1999}]{komossa_discovery_1999}
Komossa S.,  Greiner J.,  1999, åp, 349, L45

\bibitem[\protect\citeauthoryear{Kraft, Burrows  \& Nousek}{Kraft
  et~al.}{1991}]{kraft_determination_1991}
Kraft R.~P.,  Burrows D.~N.,   Nousek J.~A.,  1991, \mn@doi [The Astrophysical
  Journal] {10.1086/170124}, 374, 344

\bibitem[\protect\citeauthoryear{Krolik, Piran, Svirski  \& Cheng}{Krolik
  et~al.}{2016}]{krolik_asassn-14li_2016}
Krolik J.,  Piran T.,  Svirski G.,   Cheng R.~M.,  2016, \mn@doi [The
  Astrophysical Journal] {10.3847/0004-637X/827/2/127}, 827, 127

\bibitem[\protect\citeauthoryear{Kulkarni et~al.,}{Kulkarni
  et~al.}{2021}]{kulkarni_science_2021}
Kulkarni S.~R.,  et~al., 2021, Science with the {Ultraviolet} {Explorer}
  ({UVEX}), \url {http://arxiv.org/abs/2111.15608}

\bibitem[\protect\citeauthoryear{Leloudas et~al.,}{Leloudas
  et~al.}{2019}]{leloudas_spectral_2019}
Leloudas G.,  et~al., 2019, \mn@doi [The Astrophysical Journal]
  {10.3847/1538-4357/ab5792}, 887, 218

\bibitem[\protect\citeauthoryear{Lin, Carrasco, Grupe, Webb, Barret  \&
  Farrell}{Lin et~al.}{2011}]{lin_discovery_2011}
Lin D.,  Carrasco E.~R.,  Grupe D.,  Webb N.~A.,  Barret D.,   Farrell S.~A.,
  2011, \mn@doi [The Astrophysical Journal] {10.1088/0004-637X/738/1/52}, 738,
  52

\bibitem[\protect\citeauthoryear{Lin et~al.,}{Lin
  et~al.}{2015}]{lin_ultrasoft_2015}
Lin D.,  et~al., 2015, \mn@doi [The Astrophysical Journal]
  {10.1088/0004-637X/811/1/43}, 811, 43

\bibitem[\protect\citeauthoryear{Lin et~al.,}{Lin
  et~al.}{2017}]{lin_likely_2017}
Lin D.,  et~al., 2017, \mn@doi [Nature Astronomy] {10.1038/s41550-016-0033}, 1,
  0033

\bibitem[\protect\citeauthoryear{Lin et~al.,}{Lin
  et~al.}{2018}]{lin_luminous_2018}
Lin D.,  et~al., 2018, \mn@doi [Nature Astronomy] {10.1038/s41550-018-0493-1},
  2, 656

\bibitem[\protect\citeauthoryear{Liu, Li  \& Komossa}{Liu
  et~al.}{2014}]{liu_milliparsec_2014}
Liu F.~K.,  Li S.,   Komossa S.,  2014, \mn@doi [The Astrophysical Journal]
  {10.1088/0004-637X/786/2/103}, 786, 103

\bibitem[\protect\citeauthoryear{Liu et~al.,}{Liu
  et~al.}{2022a}]{liu_erosita_2022}
Liu T.,  et~al., 2022a, \mn@doi [Astronomy \& Astrophysics]
  {10.1051/0004-6361/202141643}, 661, A5

\bibitem[\protect\citeauthoryear{Liu, Dou, Chen  \& Shen}{Liu
  et~al.}{2022b}]{liu_uvoptical_2022}
Liu X.-L.,  Dou L.-M.,  Chen J.-H.,   Shen R.-F.,  2022b, \mn@doi [The
  Astrophysical Journal] {10.3847/1538-4357/ac33a9}, 925, 67

\bibitem[\protect\citeauthoryear{Lodato \& Rossi}{Lodato \&
  Rossi}{2011}]{lodato_multiband_2011}
Lodato G.,  Rossi E.~M.,  2011, \mn@doi [Monthly Notices of the Royal
  Astronomical Society] {10.1111/j.1365-2966.2010.17448.x}, 410, 359

\bibitem[\protect\citeauthoryear{Loeb \& Ulmer}{Loeb \&
  Ulmer}{1997}]{loeb_optical_1997}
Loeb A.,  Ulmer A.,  1997, \mn@doi [The Astrophysical Journal]
  {10.1086/304814}, 489, 573

\bibitem[\protect\citeauthoryear{Lu \& Bonnerot}{Lu \&
  Bonnerot}{2020}]{lu_self-intersection_2020}
Lu W.,  Bonnerot C.,  2020, \mn@doi [Monthly Notices of the Royal Astronomical
  Society] {10.1093/mnras/stz3405}, 492, 686

\bibitem[\protect\citeauthoryear{Lu, Beniamini  \& Bonnerot}{Lu
  et~al.}{2020}]{lu_formation_2020}
Lu W.,  Beniamini P.,   Bonnerot C.,  2020, \mn@doi [Monthly Notices of the
  Royal Astronomical Society] {10.1093/mnras/staa3372}, 500, 1817

\bibitem[\protect\citeauthoryear{Maksym, Ulmer  \& Eracleous}{Maksym
  et~al.}{2010}]{maksym_tidal_2010}
Maksym W.~P.,  Ulmer M.~P.,   Eracleous M.,  2010, \mn@doi [The Astrophysical
  Journal] {10.1088/0004-637X/722/2/1035}, 722, 1035

\bibitem[\protect\citeauthoryear{Maksym, Ulmer, Eracleous, Guennou  \&
  Ho}{Maksym et~al.}{2013}]{maksym_tidal_2013}
Maksym W.~P.,  Ulmer M.~P.,  Eracleous M.~C.,  Guennou L.,   Ho L.~C.,  2013,
  \mn@doi [Monthly Notices of the Royal Astronomical Society]
  {10.1093/mnras/stt1379}, 435, 1904

\bibitem[\protect\citeauthoryear{Malyali et~al.,}{Malyali
  et~al.}{2021}]{malyali_at_2021}
Malyali A.,  et~al., 2021, \mn@doi [Astronomy \& Astrophysics]
  {10.1051/0004-6361/202039681}, 647, A9

\bibitem[\protect\citeauthoryear{Martin et~al.,}{Martin
  et~al.}{2005}]{martin_galaxy_2005}
Martin D.~C.,  et~al., 2005, \mn@doi [The Astrophysical Journal]
  {10.1086/426387}, 619, L1

\bibitem[\protect\citeauthoryear{Masci et~al.,}{Masci
  et~al.}{2019}]{masci_zwicky_2019}
Masci F.~J.,  et~al., 2019, \mn@doi [Publications of the Astronomical Society
  of the Pacific] {10.1088/1538-3873/aae8ac}, 131, 018003

\bibitem[\protect\citeauthoryear{McConnell \& Ma}{McConnell \&
  Ma}{2013}]{mcconnell_revisiting_2013}
McConnell N.~J.,  Ma C.-P.,  2013, \mn@doi [The Astrophysical Journal]
  {10.1088/0004-637X/764/2/184}, 764, 184

\bibitem[\protect\citeauthoryear{McLure \& Dunlop}{McLure \&
  Dunlop}{2002}]{mclure_black_2002}
McLure R.~J.,  Dunlop J.~S.,  2002, \mn@doi [Monthly Notices of the Royal
  Astronomical Society] {10.1046/j.1365-8711.2002.05236.x}, 331, 795

\bibitem[\protect\citeauthoryear{Metzger \& Stone}{Metzger \&
  Stone}{2016}]{metzger_bright_2016}
Metzger B.~D.,  Stone N.~C.,  2016, \mn@doi [Monthly Notices of the Royal
  Astronomical Society] {10.1093/mnras/stw1394}, 461, 948

\bibitem[\protect\citeauthoryear{Miller}{Miller}{2015}]{miller_disk_2015}
Miller M.~C.,  2015, \mn@doi [The Astrophysical Journal]
  {10.1088/0004-637X/805/1/83}, 805, 83

\bibitem[\protect\citeauthoryear{Mockler, Guillochon  \& Ramirez-Ruiz}{Mockler
  et~al.}{2019}]{mockler_weighing_2019}
Mockler B.,  Guillochon J.,   Ramirez-Ruiz E.,  2019, \mn@doi [The
  Astrophysical Journal] {10.3847/1538-4357/ab010f}, 872, 151

\bibitem[\protect\citeauthoryear{Mummery}{Mummery}{2021}]{mummery_unified_2021}
Mummery A.,  2021, arXiv:2104.06212 [astro-ph]

\bibitem[\protect\citeauthoryear{Mummery \& Balbus}{Mummery \&
  Balbus}{2021}]{mummery_hard_2021}
Mummery A.,  Balbus S.~A.,  2021, \mn@doi [Monthly Notices of the Royal
  Astronomical Society] {10.1093/mnras/stab1184}, 504, 4730

\bibitem[\protect\citeauthoryear{Mummery \& Balbus}{Mummery \&
  Balbus}{2022}]{mummery_high-energy_2022}
Mummery A.,  Balbus S.,  2022, \mn@doi [Monthly Notices of the Royal
  Astronomical Society] {10.1093/mnras/stac2844}, 517, 3423

\bibitem[\protect\citeauthoryear{Nandra \& Pounds}{Nandra \&
  Pounds}{1994}]{nandra_ginga_1994}
Nandra K.,  Pounds K.~A.,  1994, \mn@doi [Monthly Notices of the Royal
  Astronomical Society] {10.1093/mnras/268.2.405}, 268, 405

\bibitem[\protect\citeauthoryear{Nicholl et~al.,}{Nicholl
  et~al.}{2020}]{nicholl_outflow_2020}
Nicholl M.,  et~al., 2020, \mn@doi [Monthly Notices of the Royal Astronomical
  Society] {10.1093/mnras/staa2824}, 499, 482

\bibitem[\protect\citeauthoryear{Nicholl, Lanning, Ramsden, Mockler, Lawrence,
  Short  \& Ridley}{Nicholl et~al.}{2022}]{nicholl_systematic_2022}
Nicholl M.,  Lanning D.,  Ramsden P.,  Mockler B.,  Lawrence A.,  Short P.,
  Ridley E.~J.,  2022, \mn@doi [Monthly Notices of the Royal Astronomical
  Society] {10.1093/mnras/stac2206}, 515, 5604

\bibitem[\protect\citeauthoryear{Piran, Svirski, Krolik, Cheng  \&
  Shiokawa}{Piran et~al.}{2015}]{piran_disk_2015}
Piran T.,  Svirski G.,  Krolik J.,  Cheng R.~M.,   Shiokawa H.,  2015, \mn@doi
  [The Astrophysical Journal] {10.1088/0004-637X/806/2/164}, 806, 164

\bibitem[\protect\citeauthoryear{{Planck Collaboration} et~al.,}{{Planck
  Collaboration} et~al.}{2016}]{planck_collaboration_planck_2016}
{Planck Collaboration} et~al., 2016, \mn@doi [Astronomy \& Astrophysics]
  {10.1051/0004-6361/201525830}, 594, A13

\bibitem[\protect\citeauthoryear{Predehl \& Schmitt}{Predehl \&
  Schmitt}{1995}]{predehl_x-raying_1995}
Predehl P.,  Schmitt J. H. M.~M.,  1995, Astronomy \& Astrophysics, 293, 889

\bibitem[\protect\citeauthoryear{Predehl et~al.,}{Predehl
  et~al.}{2021}]{predehl_erosita_2021}
Predehl P.,  et~al., 2021, \mn@doi [Astronomy \& Astrophysics]
  {10.1051/0004-6361/202039313}, 647, A1

\bibitem[\protect\citeauthoryear{Rees}{Rees}{1988}]{rees_tidal_1988}
Rees M.~J.,  1988, \mn@doi [Nature] {10.1038/333523a0}, 333, 523

\bibitem[\protect\citeauthoryear{Remillard et~al.,}{Remillard
  et~al.}{2022}]{remillard_empirical_2022}
Remillard R.~A.,  et~al., 2022, \mn@doi [The Astronomical Journal]
  {10.3847/1538-3881/ac4ae6}, 163, 130

\bibitem[\protect\citeauthoryear{Roming et~al.,}{Roming
  et~al.}{2005}]{roming_swift_2005}
Roming P. W.~A.,  et~al., 2005, \mn@doi [Space Science Reviews]
  {10.1007/s11214-005-5095-4}, 120, 95

\bibitem[\protect\citeauthoryear{Ryu, Krolik, Piran  \& Noble}{Ryu
  et~al.}{2020}]{ryu_tidal_2020-3}
Ryu T.,  Krolik J.,  Piran T.,   Noble S.~C.,  2020, \mn@doi [The Astrophysical
  Journal] {10.3847/1538-4357/abb3cd}, 904, 99

\bibitem[\protect\citeauthoryear{Sagiv et~al.,}{Sagiv
  et~al.}{2014}]{sagiv_science_2014}
Sagiv I.,  et~al., 2014, \mn@doi [The Astronomical Journal]
  {10.1088/0004-6256/147/4/79}, 147, 79

\bibitem[\protect\citeauthoryear{Saxton, Read, Esquej, Komossa, Dougherty,
  Rodriguez-Pascual  \& Barrado}{Saxton et~al.}{2012}]{saxton_tidal_2012}
Saxton R.~D.,  Read A.~M.,  Esquej P.,  Komossa S.,  Dougherty S.,
  Rodriguez-Pascual P.,   Barrado D.,  2012, \mn@doi [Astronomy \&
  Astrophysics] {10.1051/0004-6361/201118367}, 541, A106

\bibitem[\protect\citeauthoryear{Saxton, Read, Komossa, Lira, Alexander  \&
  Wieringa}{Saxton et~al.}{2017}]{saxton_xmmsl1_2017}
Saxton R.~D.,  Read A.~M.,  Komossa S.,  Lira P.,  Alexander K.~D.,   Wieringa
  M.~H.,  2017, \mn@doi [Astronomy \& Astrophysics]
  {10.1051/0004-6361/201629015}, 598, A29

\bibitem[\protect\citeauthoryear{Sazonov et~al.,}{Sazonov
  et~al.}{2021}]{sazonov_first_2021}
Sazonov S.,  et~al., 2021, \mn@doi [Monthly Notices of the Royal Astronomical
  Society] {10.1093/mnras/stab2843}, 508, 3820

\bibitem[\protect\citeauthoryear{Schlafly \& Finkbeiner}{Schlafly \&
  Finkbeiner}{2011}]{schlafly_measuring_2011}
Schlafly E.~F.,  Finkbeiner D.~P.,  2011, \mn@doi [The Astrophysical Journal]
  {10.1088/0004-637X/737/2/103}, 737, 103

\bibitem[\protect\citeauthoryear{Shiokawa, Krolik, Cheng, Piran  \&
  Noble}{Shiokawa et~al.}{2015}]{shiokawa_general_2015}
Shiokawa H.,  Krolik J.~H.,  Cheng R.~M.,  Piran T.,   Noble S.~C.,  2015,
  \mn@doi [The Astrophysical Journal] {10.1088/0004-637X/804/2/85}, 804, 85

\bibitem[\protect\citeauthoryear{Shu et~al.,}{Shu
  et~al.}{2020}]{shu_x-ray_2020}
Shu X.,  et~al., 2020, \mn@doi [Nature Communications]
  {10.1038/s41467-020-19675-z}, 11, 5876

\bibitem[\protect\citeauthoryear{Simard, Trevor~Mendel, Patton, Ellison  \&
  McConnachie}{Simard et~al.}{2011}]{simard_catalog_2011}
Simard L.,  Trevor~Mendel J.,  Patton D.~R.,  Ellison S.~L.,   McConnachie
  A.~W.,  2011, \mn@doi [The Astrophysical Journal Supplement Series]
  {10.1088/0067-0049/196/1/11}, 196, 11

\bibitem[\protect\citeauthoryear{Simmonds, Buchner, Salvato, Hsu  \&
  Bauer}{Simmonds et~al.}{2018}]{simmonds_xz_2018}
Simmonds C.,  Buchner J.,  Salvato M.,  Hsu L.-T.,   Bauer F.~E.,  2018,
  \mn@doi [Astronomy \& Astrophysics] {10.1051/0004-6361/201833412}, 618, A66

\bibitem[\protect\citeauthoryear{Speagle}{Speagle}{2020}]{speagle_dynesty_2020}
Speagle J.~S.,  2020, \mn@doi [Monthly Notices of the Royal Astronomical
  Society] {10.1093/mnras/staa278}, 493, 3132

\bibitem[\protect\citeauthoryear{Steinberg \& Stone}{Steinberg \&
  Stone}{2022}]{steinberg_origins_2022}
Steinberg E.,  Stone N.~C.,  2022, arXiv:2206.10641 [astro-ph]

\bibitem[\protect\citeauthoryear{Sunyaev et~al.,}{Sunyaev
  et~al.}{2021}]{sunyaev_srg_2021}
Sunyaev R.,  et~al., 2021, \mn@doi [Astronomy \& Astrophysics]
  {10.1051/0004-6361/202141179}, 656, A132

\bibitem[\protect\citeauthoryear{Torrealba, Koposov, Belokurov  \&
  Irwin}{Torrealba et~al.}{2016}]{torrealba_feeble_2016}
Torrealba G.,  Koposov S.~E.,  Belokurov V.,   Irwin M.,  2016, \mn@doi
  [Monthly Notices of the Royal Astronomical Society] {10.1093/mnras/stw733},
  459, 2370

\bibitem[\protect\citeauthoryear{Trümper}{Trümper}{1982}]{trumper_rosat_1982}
Trümper J.,  1982, \mn@doi [Advances in Space Research]
  {https://doi.org/10.1016/0273-1177(82)90070-9}, 2, 241

\bibitem[\protect\citeauthoryear{Wang et~al.,}{Wang
  et~al.}{2022}]{wang_discovery_2022}
Wang Y.,  et~al., 2022, \mn@doi [The Astrophysical Journal Letters]
  {10.3847/2041-8213/ac6670}, 930, L4

\bibitem[\protect\citeauthoryear{Weisskopf, Tananbaum, Van~Speybroeck  \&
  O'Dell}{Weisskopf et~al.}{2000}]{weisskopf_chandra_2000}
Weisskopf M.~C.,  Tananbaum H.~D.,  Van~Speybroeck L.~P.,   O'Dell S.~L.,
  2000, \mn@doi [arXiv:astro-ph/0004127] {10.1117/12.391545}, pp 2--16

\bibitem[\protect\citeauthoryear{Wen, Jonker, Stone, Zabludoff  \& Psaltis}{Wen
  et~al.}{2020}]{wen_continuum-fitting_2020}
Wen S.,  Jonker P.~G.,  Stone N.~C.,  Zabludoff A.~I.,   Psaltis D.,  2020,
  \mn@doi [The Astrophysical Journal] {10.3847/1538-4357/ab9817}, 897, 80

\bibitem[\protect\citeauthoryear{Westfall et~al.,}{Westfall
  et~al.}{2019}]{westfall_data_2019}
Westfall K.~B.,  et~al., 2019, \mn@doi [The Astronomical Journal]
  {10.3847/1538-3881/ab44a2}, 158, 231

\bibitem[\protect\citeauthoryear{Wevers et~al.,}{Wevers
  et~al.}{2019}]{wevers_evidence_2019}
Wevers T.,  et~al., 2019, \mn@doi [Monthly Notices of the Royal Astronomical
  Society] {10.1093/mnras/stz1976}, 488, 4816

\bibitem[\protect\citeauthoryear{Wevers et~al.,}{Wevers
  et~al.}{2021}]{wevers_rapid_2021}
Wevers T.,  et~al., 2021, \mn@doi [The Astrophysical Journal]
  {10.3847/1538-4357/abf5e2}, 912, 151

\bibitem[\protect\citeauthoryear{Wyrzykowski et~al.,}{Wyrzykowski
  et~al.}{2017}]{wyrzykowski_ogle16aaa_2017}
Wyrzykowski L.,  et~al., 2017, \mn@doi [Monthly Notices of the Royal
  Astronomical Society: Letters] {10.1093/mnrasl/slw213}, 465, L114

\bibitem[\protect\citeauthoryear{Yuan et~al.,}{Yuan
  et~al.}{2018}]{yuan_einstein_2018}
Yuan W.,  et~al., 2018, in den Herder J.-W.~A.,  Nakazawa K.,   Nikzad S.,
  eds, Space {Telescopes} and {Instrumentation} 2018: {Ultraviolet} to {Gamma}
  {Ray}. SPIE, Austin, United States, p.~76, \mn@doi{10.1117/12.2313358}, \url
  {https://www.spiedigitallibrary.org/conference-proceedings-of-spie/10699/2313358/Einstein-Probe--a-lobster-eye-telescope-for-monitoring-the/10.1117/12.2313358.full}

\bibitem[\protect\citeauthoryear{Zanazzi \& Ogilvie}{Zanazzi \&
  Ogilvie}{2020}]{zanazzi_eccentric_2020}
Zanazzi J.~J.,  Ogilvie G.~I.,  2020, \mn@doi [Monthly Notices of the Royal
  Astronomical Society] {10.1093/mnras/staa3127}, 499, 5562

\bibitem[\protect\citeauthoryear{van Velzen et~al.,}{van Velzen
  et~al.}{2019}]{van_velzen_first_2019}
van Velzen S.,  et~al., 2019, \mn@doi [The Astrophysical Journal]
  {10.3847/1538-4357/aafe0c}, 872, 198

\bibitem[\protect\citeauthoryear{van Velzen, Holoien, Onori, Hung  \&
  Arcavi}{van Velzen et~al.}{2020}]{van_velzen_optical-ultraviolet_2020}
van Velzen S.,  Holoien T. W.-S.,  Onori F.,  Hung T.,   Arcavi I.,  2020,
  \mn@doi [Space Science Reviews] {10.1007/s11214-020-00753-z}, 216, 124

\bibitem[\protect\citeauthoryear{van Velzen et~al.,}{van Velzen
  et~al.}{2021}]{van_velzen_seventeen_2021}
van Velzen S.,  et~al., 2021, \mn@doi [The Astrophysical Journal]
  {10.3847/1538-4357/abc258}, 908, 4

\makeatother
\end{thebibliography}




\appendix
\section{Additional material}
Table~\ref{tab:full_photometry} presents the photometric evolution of J0744, with data analysed as described in section~\ref{sec:photometric_evolution}. Table~\ref{tab:morphology} contains the results from the morphological analysis of the host galaxy of J0744.
\begin{table}
\centering
\caption{UV and optical photometry of J0744.}
\label{tab:full_photometry}
\begin{tabular}{cccc}
\hline
MJD & Instrument & Filter & Magnitude \\
\hline
59085.494 & ZTF & r & $<$20.09 \\
59085.499 & ZTF & r & $<$20.18 \\
59085.504 & ZTF & r & $<$19.99 \\
59085.509 & ZTF & r & $<$19.84 \\
59085.514 & ZTF & r & $<$19.52 \\
59087.497 & ZTF & r & $<$19.95 \\
59087.502 & ZTF & r & $<$20.10 \\
59087.507 & ZTF & r & $<$19.90 \\
59087.512 & ZTF & r & $<$19.60 \\
59089.500 & ZTF & r & $<$20.20 \\
59089.510 & ZTF & r & $<$19.68 \\
59089.515 & ZTF & r & $<$19.62 \\
59091.497 & ZTF & r & $<$20.64 \\
59091.502 & ZTF & r & $<$20.52 \\
59091.507 & ZTF & r & $<$20.60 \\
59091.512 & ZTF & r & $<$20.22 \\
59092.518 & ZTF & g & $<$19.62 \\
59092.518 & ZTF & g & $<$19.31 \\
59093.499 & ZTF & r & $<$19.88 \\
59093.504 & ZTF & r & $<$19.87 \\
59093.509 & ZTF & r & $<$20.02 \\
59093.513 & ZTF & r & $<$19.98 \\
59093.518 & ZTF & r & $<$19.69 \\
59095.498 & ZTF & r & $<$19.74 \\
59095.502 & ZTF & r & $<$19.89 \\
59095.507 & ZTF & r & $<$19.94 \\
59095.512 & ZTF & r & 19.71 $\pm$ 0.34 \\
59095.516 & ZTF & r & $<$19.76 \\
59122.509 & ZTF & r & 19.77 $\pm$ 0.17 \\
59124.406 & ZTF & r & 19.68 $\pm$ 0.34 \\
59124.476 & ZTF & r & 19.46 $\pm$ 0.17 \\
59124.526 & ZTF & g & 19.51 $\pm$ 0.23 \\
59126.498 & ZTF & g & 19.32 $\pm$ 0.18 \\
59126.524 & ZTF & r & 19.82 $\pm$ 0.22 \\
59128.445 & ZTF & g & $<$19.25 \\
59130.469 & ZTF & r & 19.89 $\pm$ 0.22 \\
59130.514 & ZTF & g & 19.56 $\pm$ 0.24 \\
59135.443 & ZTF & g & 19.53 $\pm$ 0.13 \\
59135.505 & ZTF & r & 20.02 $\pm$ 0.18 \\
59137.465 & ZTF & g & 19.81 $\pm$ 0.15 \\
59137.498 & ZTF & r & 20.03 $\pm$ 0.18 \\
59139.446 & ZTF & g & 19.55 $\pm$ 0.16 \\
59139.485 & ZTF & r & 19.93 $\pm$ 0.16 \\
59141.445 & ZTF & g & 19.83 $\pm$ 0.15 \\
59141.493 & ZTF & r & 20.03 $\pm$ 0.15 \\
59143.434 & ZTF & g & 19.89 $\pm$ 0.15 \\
59143.498 & ZTF & r & 20.30 $\pm$ 0.19 \\
59145.488 & ZTF & r & 20.16 $\pm$ 0.16 \\
59149.345 & ZTF & r & $<$19.79 \\
59149.515 & ZTF & g & 20.00 $\pm$ 0.19 \\
59150.436 & ZTF & g & 19.92 $\pm$ 0.27 \\
59150.486 & ZTF & r & 20.18 $\pm$ 0.18 \\
59152.454 & ZTF & r & 19.91 $\pm$ 0.26 \\
59152.523 & ZTF & g & 19.68 $\pm$ 0.24 \\
59155.412 & ZTF & g & 19.76 $\pm$ 0.34 \\
59155.455 & ZTF & r & 20.02 $\pm$ 0.27 \\
59157.400 & ZTF & g & $<$19.62 \\
59157.479 & ZTF & r & $<$20.00 \\
59163.440 & ZTF & g & 20.12 $\pm$ 0.27 \\
59163.503 & ZTF & r & 20.30 $\pm$ 0.22 \\
59164.854 & \textit{Swift} & UVW2 & 18.99 $\pm$ 0.12 \\
59164.857 & \textit{Swift} & UVM2 & 19.47 $\pm$ 0.16 \\
59164.860 & \textit{Swift} & UVW1 & 19.06 $\pm$ 0.12 \\
59165.422 & ZTF & r & 20.11 $\pm$ 0.19 \\
59165.444 & ZTF & g & 20.10 $\pm$ 0.19 \\
59168.436 & ZTF & r & $<$20.59 \\
59168.478 & ZTF & g & 19.98 $\pm$ 0.19 \\
59170.404 & ZTF & r & 20.65 $\pm$ 0.35 \\
59173.443 & ZTF & g & 20.11 $\pm$ 0.24 \\
59176.818 & \textit{Swift} & UVW2 & 19.34 $\pm$ 0.11 \\
59176.855 & \textit{Swift} & UVM2 & 19.43 $\pm$ 0.14 \\
59176.859 & \textit{Swift} & UVW1 & 19.34 $\pm$ 0.11 \\
59178.541 & ZTF & r & 20.26 $\pm$ 0.18 \\
59179.746 & \textit{Swift} & UVW2 & 19.37 $\pm$ 0.10 \\
59179.750 & \textit{Swift} & UVM2 & 19.53 $\pm$ 0.11 \\
59179.754 & \textit{Swift} & UVW1 & 19.61 $\pm$ 0.10 \\
59180.496 & ZTF & g & $<$20.13 \\
59182.464 & ZTF & r & 19.97 $\pm$ 0.31 \\
59182.529 & ZTF & g & $<$19.64 \\
59184.429 & ZTF & g & $<$19.78 \\
59184.464 & ZTF & r & $<$20.16 \\
59185.083 & \textit{Swift} & UVW2 & 19.51 $\pm$ 0.10 \\
59185.087 & \textit{Swift} & UVM2 & 19.60 $\pm$ 0.11 \\
59185.091 & \textit{Swift} & UVW1 & 19.54 $\pm$ 0.09 \\
59188.446 & ZTF & g & $<$19.87 \\
59193.357 & ZTF & g & 20.50 $\pm$ 0.25 \\
59193.499 & ZTF & r & 20.55 $\pm$ 0.26 \\
59195.479 & ZTF & r & 20.31 $\pm$ 0.31 \\
59198.430 & ZTF & r & $<$20.64 \\
59200.432 & ZTF & g & 19.86 $\pm$ 0.31 \\
59202.766 & \textit{Swift} & UVW2 & 19.58 $\pm$ 0.12 \\
59202.860 & \textit{Swift} & UVM2 & 19.68 $\pm$ 0.14 \\
59202.863 & \textit{Swift} & UVW1 & 19.65 $\pm$ 0.14 \\
59202.864 & \textit{Swift} & U & 19.34 $\pm$ 0.16 \\
59203.345 & ZTF & g & 20.11 $\pm$ 0.22 \\
59203.432 & ZTF & r & 20.48 $\pm$ 0.32 \\
59205.381 & ZTF & g & $<$20.59 \\
59210.264 & ZTF & g & $<$19.85 \\
59210.388 & ZTF & r & $<$20.26 \\
59212.289 & \textit{Swift} & UVW1 & 19.83 $\pm$ 0.14 \\
59212.290 & \textit{Swift} & U & 19.35 $\pm$ 0.14 \\
59212.292 & \textit{Swift} & UVW2 & 19.80 $\pm$ 0.11 \\
59216.316 & ZTF & r & $<$20.18 \\
59216.424 & ZTF & g & $<$19.09 \\
59218.266 & ZTF & g & 20.38 $\pm$ 0.36 \\
59218.403 & ZTF & r & 20.48 $\pm$ 0.34 \\
59220.298 & ZTF & r & $<$20.70 \\
59220.339 & ZTF & g & $<$20.24 \\
59220.601 & \textit{Swift} & UVM2 & 20.06 $\pm$ 0.19 \\
59220.604 & \textit{Swift} & UVW1 & 20.19 $\pm$ 0.21 \\
59220.605 & \textit{Swift} & U & 19.02 $\pm$ 0.15 \\
59220.609 & \textit{Swift} & UVW2 & 20.04 $\pm$ 0.14 \\
59222.339 & ZTF & r & 20.43 $\pm$ 0.26 \\
59222.383 & ZTF & g & 20.54 $\pm$ 0.27 \\
59224.289 & ZTF & r & $<$20.82 \\
59226.284 & ZTF & r & $<$20.66 \\
59226.387 & ZTF & g & $<$20.59 \\
59228.265 & ZTF & g & 20.70 $\pm$ 0.33 \\
59228.319 & ZTF & r & 20.49 $\pm$ 0.24 \\
59230.244 & ZTF & r & $<$20.90 \\
59230.329 & ZTF & g & $<$20.85 \\
59232.395 & ZTF & g & 20.71 $\pm$ 0.29 \\
59232.693 & \textit{Swift} & U & 19.20 $\pm$ 0.17 \\
59232.696 & \textit{Swift} & UVW2 & 20.23 $\pm$ 0.16 \\
59233.054 & \textit{Swift} & UVM2 & 20.25 $\pm$ 0.15 \\
59233.056 & \textit{Swift} & UVW1 & 20.17 $\pm$ 0.21 \\
59239.729 & \textit{Swift} & UVM2 & 20.28 $\pm$ 0.21 \\
59239.732 & \textit{Swift} & UVW1 & 20.21 $\pm$ 0.22 \\
59239.734 & \textit{Swift} & U & 18.98 $\pm$ 0.14 \\
59239.737 & \textit{Swift} & UVW2 & 20.25 $\pm$ 0.16 \\
59248.229 & ZTF & g & $<$20.83 \\
59248.237 & ZTF & r & $<$20.78 \\
59250.257 & ZTF & g & $<$20.60 \\
59250.328 & ZTF & r & $<$20.65 \\
59252.204 & ZTF & r & $<$20.99 \\
59252.287 & ZTF & g & $<$21.01 \\
59254.203 & ZTF & r & $<$20.90 \\
59254.228 & ZTF & g & $<$20.95 \\
59256.213 & ZTF & r & 20.62 $\pm$ 0.26 \\
59256.296 & ZTF & g & $<$20.78 \\
59258.483 & \textit{Swift} & UVM2 & 20.57 $\pm$ 0.31 \\
59258.484 & \textit{Swift} & UVW1 & 20.12 $\pm$ 0.26 \\
59258.485 & \textit{Swift} & U & 19.44 $\pm$ 0.24 \\
59258.485 & \textit{Swift} & UVW2 & 20.50 $\pm$ 0.22 \\
59260.193 & ZTF & g & $<$20.45 \\
59260.250 & ZTF & r & $<$20.63 \\
59263.214 & ZTF & r & $<$20.26 \\
59263.278 & ZTF & g & $<$20.41 \\
59265.188 & ZTF & g & $<$20.32 \\
59265.212 & ZTF & r & $<$20.50 \\
59267.175 & ZTF & r & $<$20.13 \\
59267.254 & ZTF & g & $<$19.79 \\
59271.225 & ZTF & g & $<$19.57 \\
59271.258 & ZTF & r & $<$19.90 \\
59273.173 & ZTF & g & $<$19.56 \\
59273.212 & ZTF & r & $<$19.76 \\
59275.172 & ZTF & g & $<$20.50 \\
59278.186 & ZTF & r & $<$20.56 \\
59280.153 & ZTF & r & 21.05 $\pm$ 0.36 \\
59280.211 & ZTF & g & $<$20.90 \\
59286.058 & \textit{Swift} & UVM2 & 20.72 $\pm$ 0.28 \\
59286.061 & \textit{Swift} & UVW1 & 21.02 $\pm$ 0.37 \\
59286.062 & \textit{Swift} & U & 19.85 $\pm$ 0.25 \\
59286.065 & \textit{Swift} & UVW2 & 20.34 $\pm$ 0.17 \\
59290.214 & ZTF & g & $<$20.93 \\
59290.239 & ZTF & r & $<$20.75 \\
59292.173 & ZTF & g & $<$20.71 \\
59292.214 & ZTF & r & $<$20.80 \\
59294.170 & ZTF & g & $<$20.34 \\
59294.208 & ZTF & r & $<$20.41 \\
59314.269 & \textit{Swift} & UVM2 & 20.91 $\pm$ 0.33 \\
59314.271 & \textit{Swift} & UVW1 & $<$20.83 \\
59314.273 & \textit{Swift} & U & 19.54 $\pm$ 0.23 \\
59314.275 & \textit{Swift} & UVW2 & 20.63 $\pm$ 0.20 \\
59342.541 & \textit{Swift} & UVM2 & $<$21.26 \\
59342.544 & \textit{Swift} & UVW1 & $<$20.83 \\
59342.546 & \textit{Swift} & U & 19.95 $\pm$ 0.32 \\
59342.549 & \textit{Swift} & UVW2 & 20.80 $\pm$ 0.20 \\
59559.674 & \textit{Swift} & UVW2 & 22.09 $\pm$ 0.33 \\
59559.678 & \textit{Swift} & UVM2 & $<$21.96 \\
59559.712 & \textit{Swift} & UVW1 & $<$20.82 \\
\end{tabular}
\end{table}

\begin{table*}
    \centering
    \begin{tabular}{c|c|c|c|c|c|c|c|c}
    \hline
         System & Model & $R_{\mathrm{HL}}$ [kpc] & $n_{\mathrm{g}}$& $n_{\mathrm{b}}$& $R_{\mathrm{d}}$ [kpc]& $B/T$ & $P_{\mathrm{pS}}$ & $P_{\mathrm{pn4}}$\\
         \hline
         SDSS~J074426.12+291607.4 (host)& S\'ersic & $\sim$2.7 & $5.88\pm 0.36$ & - & - & - & - & - \\
          & Bulge ($n_{\mathrm{b}}=4$) + Disc & $\sim 1.4$ & - & 4 & $1.3\pm 0.1$ & $0.46\pm0.03$ & 0 & - \\
          & Bulge ($n_{\mathrm{b}}=$free) + Disc & $\sim 1.8$ & - & $3.5\pm 0.3$ & $1.9\pm0.1$ & $0.48\pm0.03$ & 0 & 0 \\
         \hline 
         SDSS~J074426.49+291609.8 & S\'ersic & $\sim$4.5 & $4.39\pm 0.04$ & - & - & - & - & - \\
          & Bulge ($n_{\mathrm{b}}=4$) + Disc & $\sim 4.0$ & - & 4 & $\sim4.3$ & $\sim0.6$ & 0.04 & - \\
          & Bulge ($n_{\mathrm{b}}=$free) + Disc & $\sim 4.0$ & - & $3.78\pm 0.08$ & $\sim4.5$ & $\sim 0.6$ & 0.06 & 0.58 \\
    \end{tabular}
    \caption{Results from a morphological analysis of the pair of galaxies hosting J0744, performed in \citet{simard_catalog_2011}. $R_{\mathrm{HL}}$ is the semi-major axis half-light radius measured in the $g$-band, $n_{\mathrm{g}}$ is the galaxy S\'ersic index, $n_{\mathrm{bh}}$ is the bulge S\'ersic index, $R_{\mathrm{d}}$ is the exponential scale length of the disc, $B/T$ is the $g$-band bulge fraction. Values preceded with a $\sim$ symbol have a reported error of 0 within \citet{simard_catalog_2011}.}
    \label{tab:morphology}
\end{table*}


\bsp	
\label{lastpage}
\end{document}